\documentclass[twocolumn,10pt]{IEEEtran}

\usepackage{amsmath,amssymb,amsthm}
\usepackage{graphicx}
\usepackage[acronym]{glossaries}
\usepackage{xcolor}
\usepackage{comment}

\newacronym{mimo}{MIMO}{multiple-input multiple-output}
\newacronym{rf}{RF}{radio frequency}
\newacronym{awgn}{AWGN}{additive white Gaussian noise}
\newacronym{iid}{i.i.d.}{independent and identically distributed}
\newacronym{tx}{Tx}{transmitter}
\newacronym{rx}{Rx}{receiver}
\newacronym{milac}{MiLAC}{microwave linear analog computer}
\newacronym{snr}{SNR}{signal-to-noise ratio}
\newacronym{sim}{SIM}{stacked intelligent metasurface}
\newacronym{svd}{SVD}{singular value decomposition}
\newacronym{ris}{RIS}{reconfigurable intelligent surface}
\newacronym{bd-ris}{BD-RIS}{beyond diagonal RIS}
\newacronym{adc}{ADC}{analog-to-digital converter}
\newacronym{dac}{DAC}{digital-to-analog converter}
\newacronym{zfbf}{ZFBF}{zero-forcing beamforming}
\newacronym{mmse}{MMSE}{minimum mean square error}
\newacronym{6g}{6G}{sixth-generation}
\newacronym{5g}{5G}{fifth-generation}
\newacronym{em}{EM}{electromagnetic}

\begin{document}
\bstctlcite{BSTcontrol}

\title{Capacity of MIMO Systems Aided by\\Microwave Linear Analog Computers (MiLACs)}

\author{Matteo~Nerini,~\IEEEmembership{Member,~IEEE}, and
        Bruno~Clerckx,~\IEEEmembership{Fellow,~IEEE}

\thanks{This work has been supported in part by UKRI under Grant EP/Y004086/1, EP/X040569/1, EP/Y037197/1, EP/X04047X/1, EP/Y037243/1.}
\thanks{Matteo Nerini and Bruno Clerckx are with the Department of Electrical and Electronic Engineering, Imperial College London, SW7 2AZ London, U.K. (e-mail: m.nerini20@imperial.ac.uk; b.clerckx@imperial.ac.uk).}}

\maketitle

\begin{abstract}
Future wireless systems, known as gigantic \gls{mimo}, are expected to enhance performance by significantly increasing the number of antennas, e.g., a few thousands.
To enable gigantic \gls{mimo} overcoming the scalability limitations of digital architectures, \glspl{milac} have recently emerged.
A \gls{milac} is a multiport microwave network that processes input microwave signals entirely in the analog domain, thereby reducing hardware costs and computational complexity of gigantic \gls{mimo} architectures.
In this paper, we investigate the fundamental limits on the rate achievable in \gls{milac}-aided \gls{mimo} systems.
We model a \gls{mimo} system employing \gls{milac}-aided beamforming at the transmitter and receiver, and formulate the rate maximization problem to optimize the microwave networks of the \glspl{milac}, which are assumed lossless and reciprocal for practical reasons.
Under the lossless and reciprocal constraints, we derive a global optimal solution for the microwave networks of the \glspl{milac} in closed form.
In addition, we also characterize in closed-form the capacity of \gls{mimo} systems operating \gls{milac}-aided beamforming.
Our theoretical analysis, confirmed by numerical simulations, reveals that \gls{milac}-aided beamforming achieves the same capacity as digital beamforming, while significantly reducing the number of \gls{rf} chains, \glspl{adc}/\glspl{dac} resolution requirements, and computational complexity.
\end{abstract}

\glsresetall


\section{Introduction}

\Gls{6g} wireless networks aim to deliver unprecedented data rates, ultra-low latency, and highly reliable connectivity.
While massive \gls{mimo} technology, typically employing 64 antennas, has been effective for \gls{5g} networks \cite{lar14,bjo16}, further scaling is essential to meet the requirements of \gls{6g} applications in the envisioned upper mid-band frequencies (7-24 GHz).
For this reason, gigantic \gls{mimo} has been introduced as the new \gls{6g} \gls{mimo} technology \cite{bjo24}, denoted as Giga-\gls{mimo} in the Qualcomm \gls{6g} Vision White Paper \cite{qua22}.
By scaling the number of antennas dramatically up to a few thousands, gigantic \gls{mimo} is expected to enable finer beamforming and enhance spatial multiplexing, essential in \gls{6g} networks.

\begin{figure}[t]
\centering
\includegraphics[width=0.36\textwidth]{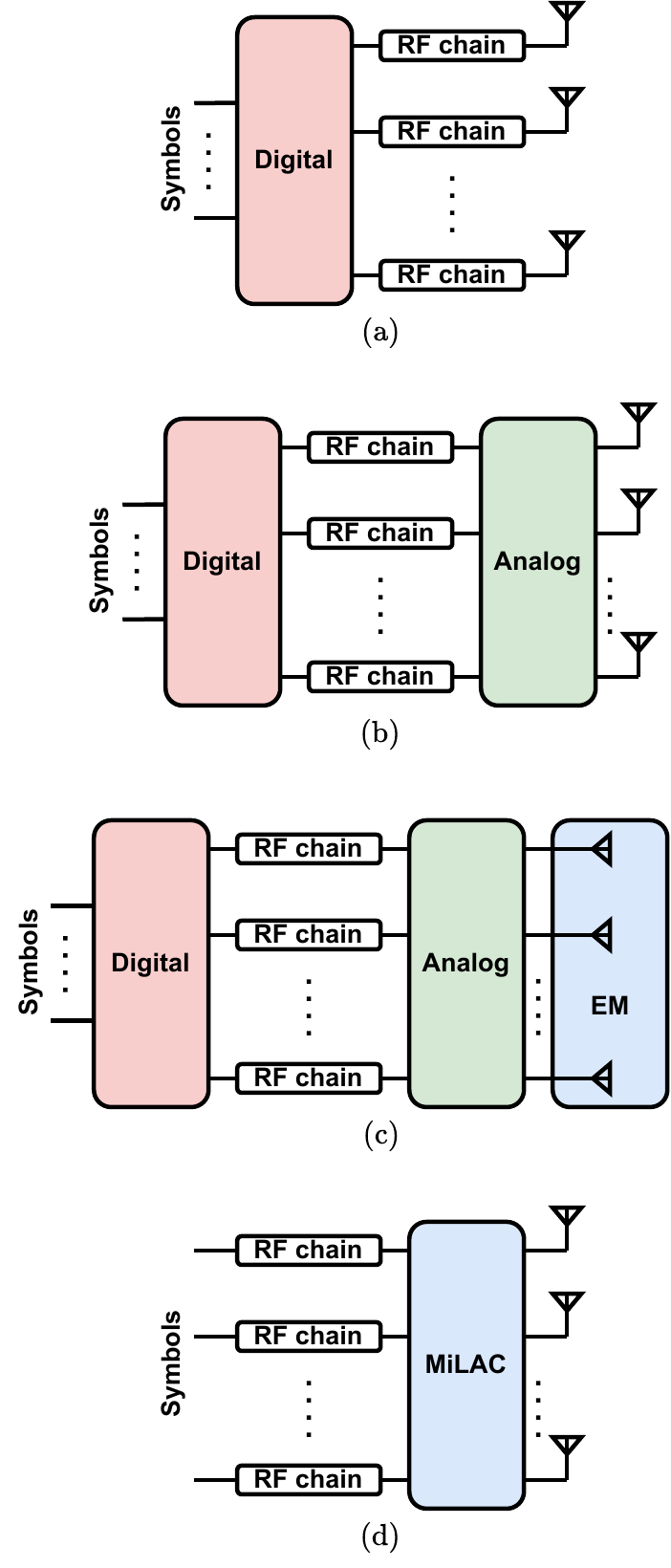}
\caption{Representation of (a) digital, (b) hybrid, (c) tri-hybrid, and (d) MiLAC-aided MIMO architectures.}
\label{fig:comparison}
\end{figure}

Over the past decade, significant research efforts have been dedicated to designing \gls{mimo} architectures scalable to high numbers of antennas.
Conventionally, fully digital \gls{mimo} architectures precode and combine the symbols in the digital domain, and require a \gls{rf} chain connected to each antenna element, as shown in Fig.~\ref {fig:comparison}(a).
Since the symbols are processed digitally, digital \gls{mimo} architectures offer maximum flexibility and, as a consequence, maximum performance.
However, the \gls{rf} chains include expensive and power-hungry components such as high-resolution \glspl{adc}/\glspl{dac} and mixers, making this solution not scalable to gigantic \gls{mimo}.
For this reason, several solutions have been proposed to perform beamforming partially or fully in the analog domain, easing the requirements on the number of \gls{rf} chains and the resolution of \glspl{adc}/\glspl{dac}.

A popular alternative to digital beamforming is hybrid beamforming, where the symbols are precoded in the digital (or baseband) domain and also in the analog (or \gls{rf}) domain through a network of phase shifters or time delay units \cite{sun14,aya14,soh16}, as shown in Fig.~\ref{fig:comparison}(b).
By processing the symbols also in the analog domain, hybrid beamforming can achieve performance comparable to that of digital beamforming, even with a reduced number of \gls{rf} chains, which can be much lower than the number of antennas.
Very recently, to further reduce the power consumption without sacrificing performance, the hybrid \gls{mimo} architecture has been generalized into the so-called tri-hybrid \gls{mimo} architecture \cite{cas25}, represented in Fig.~\ref{fig:comparison}(c).
In tri-hybrid \gls{mimo}, the symbols are processed in three stages: \textit{i)} in the digital domain, \textit{ii)} in the analog domain through phase shifters or time delay units as in hybrid beamforming, and \textit{iii)} in the \gls{em} domain through reconfigurable antennas.
Examples of reconfigurable antenna technologies that can be used in tri-hybrid beamforming include metasurface antennas or \gls{ris} \cite{wu21}, parasitic element-assisted antennas \cite{kaw05}, and fluid or movable antennas \cite{won21,zhu24}.

Exploiting reconfigurable antenna technologies, several strategies have been proposed to perform beamforming directly in the \gls{em}-domain.
\glspl{ris} deployed close to a transmitting device have been considered to steer the radiated signal toward the intended receiver \cite{jam21,you22,hua23}.
To improve the flexibility of \gls{ris}, \gls{bd-ris} has emerged as a general family of \gls{ris} architectures allowing interconnections between the elements \cite{she22,li24,ner24-1}, and \gls{sim} has been proposed by exploiting the flexibility of multiple stacked \glspl{ris} \cite{an23,an24,ner24-3}.
In addition, dynamic scattering arrays have been studied, where multiple scattering elements jointly contribute to the \gls{em} processing of the radiated signal \cite{dar24}.
The use of impedance networks operating in the \gls{em} domain has also been considered to reduce the burden on the baseband unit in decentralized multi-antenna architectures \cite{ale24}.

Following this trend of shifting from digital to analog or \gls{em} processing, the concept of \gls{milac} has recently been proposed in \cite{ner25-1,ner25-2}.
A \gls{milac} is a multiport microwave network composed of tunable impedance components, designed to receive input signals on specific ports and deliver output signals processed entirely in the analog domain.
By exploiting the massive parallelism offered by analog computing, a \gls{milac} can compute various operations with remarkably reduced computational complexity, achieving performance levels unattainable with conventional digital computing \cite{ner25-1}.
One notable application of \gls{milac} is in wireless communications, where \gls{milac}-aided beamforming has been introduced to enable efficient beamforming in gigantic \gls{mimo} systems \cite{ner25-2}.
\Gls{milac}-aided beamforming is illustrated in Fig.~\ref{fig:comparison}(d), showing that some of the ports of the \gls{milac} are connected to the \gls{rf} chains carrying the symbols, while others are connected to the antennas.

Since \gls{milac}-aided beamforming processes the transmitted symbols fully in the analog domain, it offers five advantages that significantly reduce hardware cost and computational complexity, thereby enabling gigantic \gls{mimo} \cite{ner25-2}.
\textit{First}, it can achieve the same flexibility and performance as digital beamforming.
\textit{Second}, it requires only the minimum number of \gls{rf} chains, equal to the number of transmitted symbols (or streams).
\textit{Third}, it allows the use of low-resolution \glspl{adc}/\glspl{dac} because the \gls{rf} chains carry the symbols rather than their linear combinations.
\textit{Fourth}, it precodes and combines the symbols directly in the analog domain, thus avoiding the need for a matrix-vector multiplication at every symbol time.
\textit{Fifth}, it can perform \gls{zfbf} at the transmitter and \gls{mmse} detection at the receiver with remarkably reduced computational complexity.

To investigate the fundamental limits of the capabilities of a \gls{milac}, \cite{ner25-2} has considered \glspl{milac} having microwave networks that can be arbitrarily reconfigured.
However, this design presents two main practical challenges.
First, in \cite{ner25-2} it is assumed that the tunable impedance components of the \gls{milac} can be reconfigured to any complex value.
In practice, it is more feasible to reconfigure only the imaginary part of the impedance components and assume that the real part is approximately zero, i.e., to consider lossless impedance components.
This constraint ensures that no power supplies are needed and no power is dissipated within the \gls{milac}.
Second, the design in \cite{ner25-2} allows the impedance components of the \gls{milac} to be non-reciprocal, meaning that their impedance values depend on the direction of propagation of the signal.
However, to avoid the need for non-reciprocal \gls{rf} components, it is more practical to employ only reciprocal impedance components, which maintain the same value regardless of the direction of the signal.
Although it has been shown that a lossless \gls{milac} can achieve the same performance benefits as an arbitrarily reconfigurable one \cite{ner25-1}, the performance limits of \glspl{milac} that are both lossless and reciprocal remain unknown.
To address this gap, in this study, we focus on \glspl{milac} that are constrained to be both lossless and reciprocal.
Specifically, we investigate their optimization and characterize the fundamental limits on the rate achievable by \gls{milac}-aided beamforming using lossless and reciprocal \glspl{milac}.
The main contributions of this paper are summarized as follows.

\begin{figure*}[t]
\centering
\includegraphics[height=0.268\textwidth]{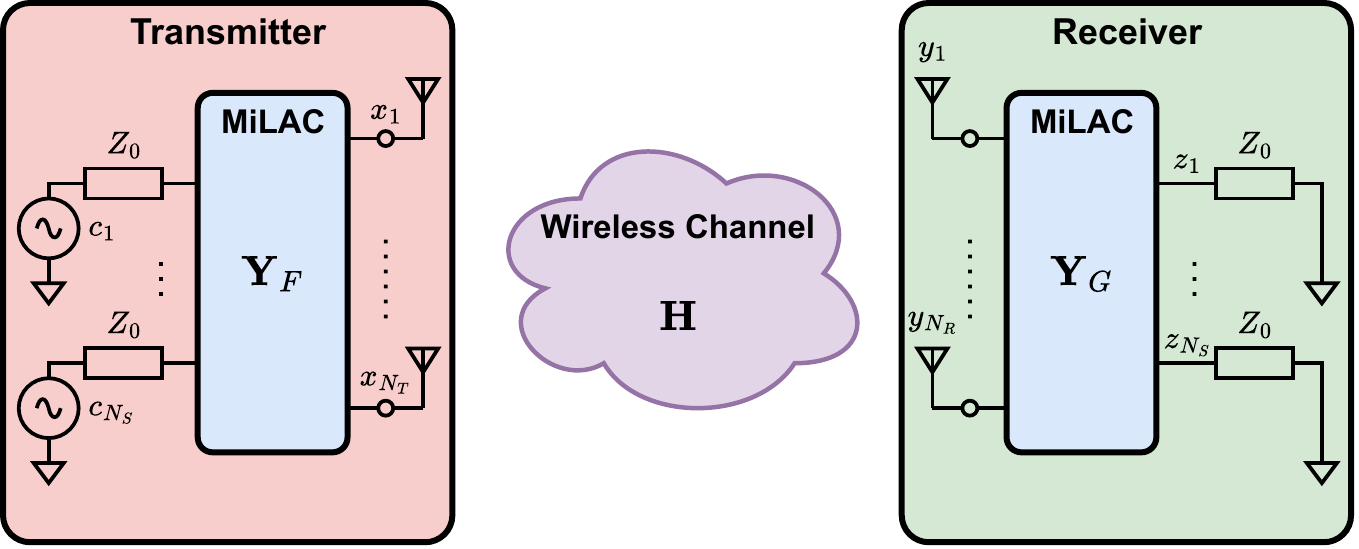}
\caption{MIMO system with MiLAC at both the transmitter and receiver.}
\label{fig:milac}
\end{figure*}

\textit{First}, we model a point-to-point \gls{mimo} system where both receiver and transmitter are equipped with a \gls{milac} to perform \gls{milac}-aided beamforming, as represented in Fig.~\ref{fig:milac}.
For practical reasons, we consider lossless and reciprocal \glspl{milac}, modeling the corresponding constraints of their microwave networks.
Based on the obtained \gls{milac}-aided \gls{mimo} system model, we formulate the rate maximization problem accounting for the fact that the symbols are precoded at the transmitter and recovered at the receiver only operating in the analog domain through the \glspl{milac}.

\textit{Second}, we solve the rate maximization problem with a closed-form global optimal solution, i.e., capacity-achieving.
To this end, we first model the \glspl{milac} through scattering parameters, which allow us to simplify the constraints.
The problem is further simplified and solved in closed form by maximizing an upper bound on the rate and relaxing its constraints.
We then prove that the derived closed-form solution globally maximizes the rate and fulfills the constraints of the original optimization problem.
The capacity of \gls{milac}-aided \gls{mimo} systems is also characterized in closed form.

\textit{Third}, we rigorously compare the capacity achieved by \gls{milac}-aided beamforming with the capacity of an equivalent point-to-point \gls{mimo} system operating with digital beamforming.
Analytical results show that lossless and reciprocal \glspl{milac} can achieve the same capacity as digital beamforming with the same number of streams.
This makes \gls{milac}-aided beamforming an appealing solution to efficiently enable gigantic \gls{mimo} communications with no performance degradation compared to digital beamforming.

\textit{Organization}:
In Section~\ref{sec:system}, we model a \gls{mimo} system operating \gls{milac}-aided beamforming at both the transmitter and receiver.
In Section~\ref{sec:problem}, we formulate the rate maximization problem for a \gls{mimo} system with \gls{milac}-aided beamforming.
In Section~\ref{sec:solution}, we solve the rate maximization problem by maximizing an upper bound on the rate and relaxing its constraints.
In Section~\ref{sec:optimal}, we show that the derived solution globally maximizes the rate and fulfills the original constraints.
In Section~\ref{sec:digital}, we show that \gls{milac}-aided beamforming can achieve the same capacity as digital beamforming.
In Section~\ref{sec:results}, we validate our theoretical findings through numerical results.
Finally, Section~\ref{sec:conclusion} concludes this paper.

\textit{Notation}:
Vectors and matrices are denoted with bold lower and bold upper letters, respectively.
Scalars are represented with letters not in bold font.
$\Re\{a\}$, $\Im\{a\}$, and $\vert a\vert$ refer to the real part, imaginary part, and absolute value of a complex scalar $a$, respectively.
$\mathbf{a}^*$, $\mathbf{a}^T$, $\mathbf{a}^H$, $[\mathbf{a}]_{i}$, and $\Vert\mathbf{a}\Vert$ refer to the conjugate, transpose, conjugate transpose, $i$th element, and $l_{2}$-norm of a vector $\mathbf{a}$, respectively.
$\mathbf{A}^*$, $\mathbf{A}^T$, $\mathbf{A}^H$, $[\mathbf{A}]_{i,k}$, $[\mathbf{A}]_{i,:}$, $[\mathbf{A}]_{:,k}$, and $\Vert\mathbf{A}\Vert_F$ refer to the conjugate, transpose, conjugate transpose, $(i,k)$th element, $i$th row, $k$th column, and Frobenius norm of a matrix $\mathbf{A}$, respectively.
$[\mathbf{A}]_{\mathcal{I},\mathcal{K}}$ refers to the submatrix of $\mathbf{A}$ obtained by selecting the rows and columns indexed by the elements of the sets $\mathcal{I}$ and $\mathcal{K}$, respectively.
$\mathbb{R}$ and $\mathbb{C}$ denote the real and complex number sets, respectively.
$j=\sqrt{-1}$ denotes the imaginary unit.
$\mathbf{I}_N$ and $\mathbf{0}_N$ denote the identity matrix and the all-zero matrix with dimensions $N\times N$, respectively.
$\mathbf{0}_{M\times N}$ denotes the all-zero matrix with dimensions $M\times N$.

\section{Microwave Linear Analog Computer (MiLAC)-Aided MIMO System Model}
\label{sec:system}

In this section, we introduce the system model of a point-to-point \gls{mimo} system where the transmitter and receiver are equipped with a \gls{milac}.
In addition, we characterize a \gls{milac} implemented with lossless and reciprocal admittance components and derive the resulting constraints on the admittance matrix of its microwave network.

\subsection{System Model}

Consider a point-to-point \gls{mimo} system aided by a \gls{milac} at both the transmitter and receiver, as represented in Fig.~\ref{fig:milac}.
The transmitter, equipped with $N_T$ antennas, sends $N_S$ symbols in parallel, also known as streams, to the receiver, which has $N_R$ antennas, where $N_S\leq\min\{N_T,N_R\}$.
As both transmitter and receiver operate \gls{milac}-aided beamforming, they both include only $N_S$ \gls{rf} chains \cite{ner25-2}, which is the minimum number of \gls{rf} chains needed to exchange $N_S$ streams in parallel.
In Fig.~\ref{fig:milac}, the \gls{rf} chains at the transmitter are represented through voltage generators with their series impedance $Z_0$, commonly set to $Z_0=50~\Omega$, while the \gls{rf} chains at the receiver are represented through terminals loaded with the reference impedance $Z_0$ \cite{ivr10,ner24-2}.

We denote the symbol vector as $\mathbf{s}\in\mathbb{C}^{N_S\times1}$, which contains the $N_S$ symbols sent by the transmitter such that $\mathbb{E}[\mathbf{s}\mathbf{s}^H]=\mathbf{I}_{N_S}$.
The source signal at the transmitting \gls{rf} chains is denoted as $\mathbf{c}\in\mathbb{C}^{N_S\times1}$ and contains the $N_S$ transmitted symbols with their allocated power, i.e.,
\begin{equation}
\mathbf{c}=\sqrt{P_T}\mathbf{P}^{1/2}\mathbf{s},\label{eq:c}
\end{equation}
where $\mathbf{P}^{1/2}\in\mathbb{C}^{N_S\times N_S}$ is the square root of the power allocation matrix given by $\mathbf{P}^{1/2}=\text{diag}(\sqrt{p_1},\ldots,\sqrt{p_{N_S}})$, with $p_{s}$ denoting the power allocation for the $s$th symbol such that $\sum_{s=1}^{N_S}p_{s}=1$, and $P_T$ is the transmitted power.
In Fig.~\ref{fig:milac}, the source signal $\mathbf{c}$ is the signal imposed by the voltage generators.
Specifically, the voltage at the $s$th generator is $c_s$, for $s=1,\ldots,N_S$, with $\mathbf{c}=[c_1,\ldots,c_{N_S}]^T$.
Thus, the average power of the signal at the voltage generators $\mathbf{c}$ is $P_T$, as it can be shown by computing $\mathbb{E}[\Vert\mathbf{c}\Vert^2]=\mathbb{E}[\Vert\sqrt{P_T}\mathbf{P}^{1/2}\mathbf{s}\Vert^2]=P_T\mathbb{E}[\text{Tr}(\mathbf{P}^{1/2}\mathbf{s}\mathbf{s}^H\mathbf{P}^{1/2})]=P_T\text{Tr}(\mathbf{P})=P_T$.
The signal at the \gls{rf} chains $\mathbf{c}$ is precoded by a \gls{milac}, giving the signal at the $N_T$ transmitting antennas $\mathbf{x}\in\mathbb{C}^{N_T\times1}$ as
\begin{equation}
\mathbf{x}=\mathbf{F}\mathbf{c},\label{eq:x}
\end{equation}
where $\mathbf{F}\in\mathbb{C}^{N_T\times N_S}$ is the precoding matrix implemented by the \gls{milac} at the transmitter given by
\begin{equation}
\mathbf{F}=\left[\left(\frac{\mathbf{Y}_F}{Y_0}+\mathbf{I}_{N_S+N_T}\right)^{-1}\right]_{N_S+1:N_S+N_T,1:N_S},\label{eq:F(Y)1}
\end{equation}
as a function of the admittance matrix of the \gls{milac} at the transmitter $\mathbf{Y}_F\in\mathbb{C}^{(N_S+N_T)\times(N_S+N_T)}$ \cite[Chapter~4]{poz11}, \cite{ner25-2}.
To simplify the notation in the remainder of the paper, we introduce the two sets $\mathcal{N}_T=\{1,\ldots,N_T\}$ and $\mathcal{N}_S=\{1,\ldots,N_S\}$, such that \eqref{eq:F(Y)1} can be rewritten in a more compact form as
\begin{equation}
\mathbf{F}=\left[\left(\frac{\mathbf{Y}_F}{Y_0}+\mathbf{I}_{N_S+N_T}\right)^{-1}\right]_{N_S+\mathcal{N}_T,\mathcal{N}_S}.\label{eq:F(Y)2}
\end{equation}

The signal at the $N_R$ receiving antennas is denoted as $\mathbf{y}\in\mathbb{C}^{N_R\times1}$, and writes as
\begin{equation}
\mathbf{y}=\mathbf{H}\mathbf{x}+\mathbf{n},\label{eq:y}
\end{equation}
where $\mathbf{H}\in\mathbb{C}^{N_R\times N_T}$ is the wireless channel between the transmitter and receiver and $\mathbf{n}\in\mathbb{C}^{N_R\times 1}$ is the \gls{awgn} such that $\mathbb{E}[\mathbf{n}\mathbf{n}^H]=\sigma^2\mathbf{I}_{N_R}$, with $\sigma^2$ denoting the noise power.
Note that the expression in \eqref{eq:y} is the baseband representation of the received signal $\mathbf{y}$, assuming that it is sampled at the symbol rate.
At the receiver, the signal $\mathbf{y}$ is combined by a \gls{milac} to obtain the signal used for detection $\mathbf{z}\in\mathbb{C}^{N_S\times 1}$ as
\begin{equation}
\mathbf{z}=\mathbf{G}\mathbf{y},\label{eq:z}
\end{equation}
where $\mathbf{G}\in\mathbb{C}^{N_S\times N_R}$ is the combining matrix implemented by the \gls{milac} written as
\begin{equation}
\mathbf{G}=\left[\left(\frac{\mathbf{Y}_G}{Y_0}+\mathbf{I}_{N_R+N_S}\right)^{-1}\right]_{N_R+\mathcal{N}_S,\mathcal{N}_R},\label{eq:G(Y)}
\end{equation}
as a function of the admittance matrix of the \gls{milac} at the receiver $\mathbf{Y}_G\in\mathbb{C}^{(N_R+N_S)\times(N_R+N_S)}$ \cite[Chapter~4]{poz11} \cite{ner25-2}, where we have introduced the set $\mathcal{N}_R=\{1,\ldots,N_R\}$ to simplify the notation.
In Fig.~\ref{fig:milac}, $\mathbf{z}$ is the signal visible at the receiving \gls{rf} chains.
Specifically, the voltage at the $s$th \gls{rf} chain is $z_s$, for $s=1,\ldots,N_S$, with $\mathbf{z}=[z_1,\ldots,z_{N_S}]^T$.
By substituting \eqref{eq:c}, \eqref{eq:x}, and \eqref{eq:y} into \eqref{eq:z}, the end-to-end system model is given by
\begin{equation}
\mathbf{z}=\sqrt{P_T}\mathbf{G}\mathbf{H}\mathbf{F}\mathbf{P}^{1/2}\mathbf{s}+\mathbf{G}\mathbf{n},\label{eq:system-model}
\end{equation}
relating the signal used for detection $\mathbf{z}$ to the transmitted symbols $\mathbf{s}$.

\subsection{MiLACs}

In \cite{ner25-1,ner25-2}, it has been shown that a \gls{milac} with no constraints on its admittance matrix can implement any arbitrary beamforming matrix, thereby achieving the same flexibility and performance as digital beamforming.
In such a \gls{milac}, the ports are interconnected to ground and to each other through tunable admittance components allowed to assume any complex value.
Considering the \gls{milac} at the transmitter, port $k$ is interconnected to ground through an admittance $Y_{F,k,k}\in\mathbb{C}$, for $k=1,\ldots,N_S+N_T$, and it is interconnected to port $i$ through an admittance $Y_{F,i,k}\in\mathbb{C}$, $\forall i\neq k$.
As a function of these tunable admittance components, the entries of the admittance matrix of the \gls{milac} are given by
\begin{equation}
\left[\mathbf{Y}_F\right]_{i,k}=
\begin{cases}
-Y_{F,i,k} & i\neq k\\
\sum_{n=1}^{N_S+N_T}Y_{F,n,k} & i=k
\end{cases},\label{eq:Yik-tx}
\end{equation}
for $i,k=1,\ldots,N_S+N_T$, as discussed in \cite{ner25-1,ner25-2}.
Similarly, in the \gls{milac} at the receiver side, port $k$ is interconnected to ground through an admittance $Y_{G,k,k}\in\mathbb{C}$, for $k=1,\ldots,N_R+N_S$, and it is interconnected to port $i$ through an admittance $Y_{G,i,k}\in\mathbb{C}$, $\forall i\neq k$.
The admittance matrix of the \gls{milac} at the receiver is therefore given by
\begin{equation}
\left[\mathbf{Y}_G\right]_{i,k}=
\begin{cases}
-Y_{G,i,k} & i\neq k\\
\sum_{n=1}^{N_R+N_S}Y_{G,n,k} & i=k
\end{cases},\label{eq:Yik-rx}
\end{equation}
for $i,k=1,\ldots,N_R+N_S$.

When the admittance components of a \gls{milac} can be tuned to any complex value, the admittance matrix of the \gls{milac} can be arbitrarily reconfigured, and any beamforming matrix can be implemented.
However, implementing such a \gls{milac} poses two practical problems.
First, admittance components with a non-zero real part are either implemented through active components or resistive loads, both increasing the power consumption of the \gls{milac}.
Second, to implement a \gls{milac} with non-reciprocal components poses additional challenges since common admittance components such as resistors, inductors, and capacitors are reciprocal components, i.e., their value is independent of the direction of propagation of the signal.
To overcome these two problems, we consider in this work \glspl{milac} that are solely composed of lossless and reciprocal admittance components.

\subsection{Lossless and Reciprocal MiLACs}

\begin{figure}[t]
\centering
\includegraphics[width=0.42\textwidth]{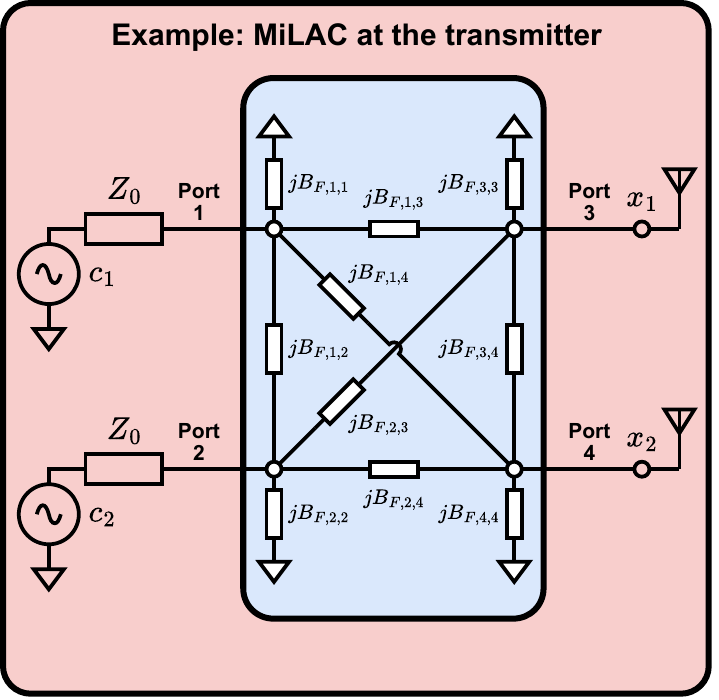}
\caption{Example of lossless and reciprocal MiLAC at the transmitter, with $N_S=2$ streams and $N_T=2$ antennas.}
\label{fig:tx}
\end{figure}

When the \glspl{milac} at the transmitter and receiver are composed of lossless and reciprocal admittance components, all their admittance components are purely imaginary, and their value is the same regardless of the direction of propagation of the signal.
In the case of a lossless \gls{milac} at the transmitter, its purely imaginary admittance components can be expressed as $Y_{F,i,k}=jB_{F,i,k}$, where $B_{F,i,k}\in\mathbb{R}$ is the susceptance value of $Y_{F,i,k}$, for $i,k=1,\ldots,N_S+N_T$.
Thus, by substituting $Y_{F,i,k}=jB_{F,i,k}$ into \eqref{eq:Yik-tx}, we obtain that the admittance matrix of the \gls{milac} is purely imaginary.
Besides, when the \gls{milac} at the transmitter is reciprocal, we have $Y_{F,i,k}=Y_{F,k,i}$, $\forall i\neq k$, imposing its admittance matrix to be symmetric because of \eqref{eq:Yik-tx}.
For a lossless and reciprocal \gls{milac} at the transmitter, when these two constraints are jointly considered, it holds
\begin{equation}
\mathbf{Y}_F=j\mathbf{B}_F,\;\mathbf{B}_F=\mathbf{B}_F^T,
\end{equation}
where $\mathbf{B}_F\in\mathbb{R}^{(N_S+N_T)\times(N_S+N_T)}$ is the susceptance matrix of the \gls{milac}.
Note that reconfigurable microwave networks with the same constraints have been proposed to implement \gls{bd-ris} to shape the wireless channel with enhanced flexibility compared to conventional \gls{ris} \cite{she22,ner24-1}.
For the sake of example, the circuit of a lossless and reciprocal MiLAC at the transmitter sending $N_S=2$ streams over $N_T=2$ antennas is illustrated in Fig.~\ref{fig:tx}.

\begin{figure}[t]
\centering
\includegraphics[width=0.42\textwidth]{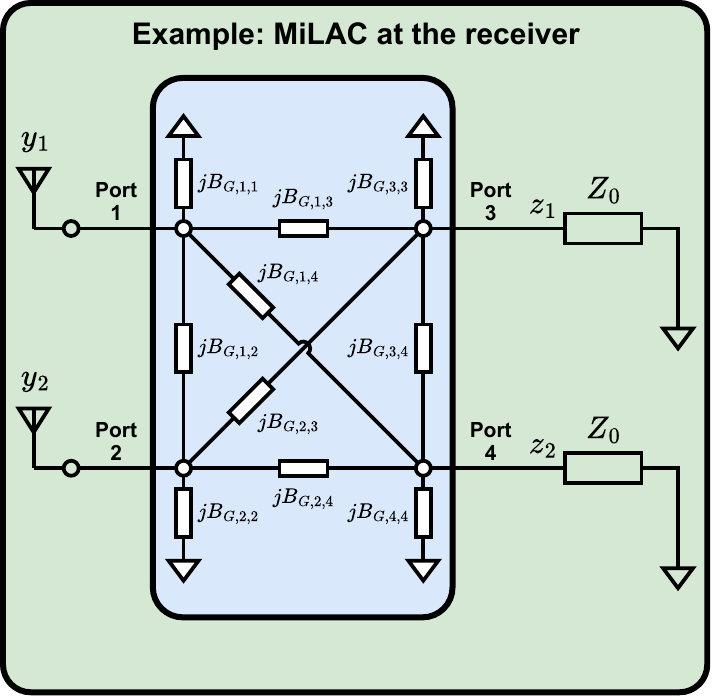}
\caption{Example of lossless and reciprocal MiLAC at the receiver, with $N_R=2$ antennas and $N_S=2$ streams.}
\label{fig:rx}
\end{figure}

Following a similar discussion at the receiver side, in a lossless \gls{milac} at the receiver the admittance components read as $Y_{G,i,k}=jB_{G,i,k}$, where $B_{G,i,k}\in\mathbb{R}$ is the susceptance value of $Y_{G,i,k}$, for $i,k=1,\ldots,N_R+N_S$.
Furthermore, if the \gls{milac} at the receiver is reciprocal it holds $Y_{G,i,k}=Y_{G,k,i}$, $\forall i\neq k$.
By jointly considering these two properties, the admittance matrix of a lossless and reciprocal \gls{milac} at the receiver is constrained as
\begin{equation}
\mathbf{Y}_G=j\mathbf{B}_G,\;\mathbf{B}_G=\mathbf{B}_G^T,
\end{equation}
with $\mathbf{B}_G\in\mathbb{R}^{(N_R+N_S)\times(N_R+N_S)}$ being the susceptance matrix of the \gls{milac}.
As an example, the circuit of a lossless and reciprocal MiLAC at the receiver with $N_R=2$ antennas and $N_S=2$ streams is illustrated in Fig.~\ref{fig:rx}.

\section{Rate Maximization Problem}
\label{sec:problem}

In this section, we formulate the rate maximization problem for a point-to-point \gls{mimo} system where the transmitter and receiver are equipped with lossless and reciprocal \glspl{milac}.

The goal of \gls{milac}-aided beamforming is to operate precoding and combining entirely in the analog domain, at the transmitter and receiver, respectively, reducing the resolution needed at the \glspl{adc} and \glspl{dac} and the required computational complexity.
Considering this limitation, the receiver detects the symbols $\mathbf{s}$ directly from the signal $\mathbf{z}$, and avoids further combining of the signal $\mathbf{z}$ in the digital domain.
Thus, recalling the system model in \eqref{eq:system-model}, the achievable rate of our \gls{milac}-aided \gls{mimo} system is defined as
\begin{equation}
R=\sum_{s=1}^{N_S}\log_2\left(1+\frac{P_Tp_{s}\left\vert[\mathbf{G}\mathbf{H}\mathbf{F}]_{s,s}\right\vert^2}
{P_T\sum_{t\neq s}p_{t}\left\vert[\mathbf{G}\mathbf{H}\mathbf{F}]_{s,t}\right\vert^2+\left\Vert[\mathbf{G}]_{s,:}\right\Vert^2\sigma^2}\right),\label{eq:C1}
\end{equation}
where for the $s$th stream the interference from the other symbols is treated as noise, for $s=1,\ldots,N_S$, and we have assumed Gaussian input signal $\mathbf{s}$.\footnote{In the following, the ``achievable rate'' is referred to as the ``rate'' for brevity.
Note that a similar definition of rate has been adopted for \gls{sim}-aided \gls{mimo} systems following the same motivation \cite{an23}.}
Denoting as $\mathbf{f}_s\in\mathbb{C}^{N_T\times1}$ the $s$th column of $\mathbf{F}$, for $s=1,\ldots,N_S$, i.e., $\mathbf{F}=[\mathbf{f}_1,\ldots,\mathbf{f}_{N_S}]$, and as $\mathbf{g}_s\in\mathbb{C}^{1\times N_R}$ the $s$th row of $\mathbf{G}$, for $s=1,\ldots,N_S$, i.e., $\mathbf{G}=[\mathbf{g}_1^T,\ldots,\mathbf{g}_{N_S}^T]^T$, we can rewrite \eqref{eq:C1} as
\begin{equation}
R=\sum_{s=1}^{N_S}\log_2\left(1+\frac{P_Tp_{s}\left\vert\mathbf{g}_s\mathbf{H}\mathbf{f}_s\right\vert^2}
{P_T\sum_{t\neq s}p_{t}\left\vert\mathbf{g}_s\mathbf{H}\mathbf{f}_t\right\vert^2+\left\Vert\mathbf{g}_{s}\right\Vert^2\sigma^2}\right),\label{eq:C2}
\end{equation}
and, introducing $\hat{\mathbf{g}}_s=\mathbf{g}_s/\Vert\mathbf{g}_s\Vert$, for $s=1,\ldots,N_S$, \eqref{eq:C2} can be simplified as
\begin{equation}
R=\sum_{s=1}^{N_S}\log_2\left(1+\frac{P_Tp_{s}\left\vert\hat{\mathbf{g}}_s\mathbf{H}\mathbf{f}_s\right\vert^2}
{P_T\sum_{t\neq s}p_{t}\left\vert\hat{\mathbf{g}}_s\mathbf{H}\mathbf{f}_t\right\vert^2+\sigma^2}\right),\label{eq:C3}
\end{equation}
showing that the rate is independent of the $\ell_2$-norms of the rows of the combining matrix $\mathbf{G}$.

Given the achievable rate expressed as in \eqref{eq:C3}, the rate maximization problem is formalized as
\begin{align}
\underset{\mathbf{B}_F,\mathbf{B}_G,p_{s}}{\mathsf{\mathrm{max}}}\;\;
&R\label{eq:prob1-ini}\\
\mathsf{\mathrm{s.t.}}\;\;\;
&\mathbf{F}=\left[\left(\frac{\mathbf{Y}_F}{Y_0}+\mathbf{I}_{N_S+N_T}\right)^{-1}\right]_{N_S+\mathcal{N}_T,\mathcal{N}_S},\label{eq:prob1-c1}\\
&\mathbf{Y}_F=j\mathbf{B}_F,\;\mathbf{B}_F=\mathbf{B}_F^T,\label{eq:prob1-c2}\\
&\mathbf{G}=\left[\left(\frac{\mathbf{Y}_G}{Y_0}+\mathbf{I}_{N_R+N_S}\right)^{-1}\right]_{N_R+\mathcal{N}_S,\mathcal{N}_R},\label{eq:prob1-c3}\\
&\mathbf{Y}_G=j\mathbf{B}_G,\;\mathbf{B}_G=\mathbf{B}_G^T,\label{eq:prob1-c4}\\
&\mathbf{P}=\text{diag}(p_1,\ldots,p_{N_S}),\;\sum_{s=1}^{N_S}p_{s}=1,\label{eq:prob1-fin}
\end{align}
where constraints \eqref{eq:prob1-c1} and \eqref{eq:prob1-c2} express how the precoding matrix $\mathbf{F}$ is constrained in the case of a lossless and reciprocal \gls{milac} at the transmitter, \eqref{eq:prob1-c3} and \eqref{eq:prob1-c4} characterize the combining matrix $\mathbf{G}$ in the case of a lossless and reciprocal \gls{milac} at the receiver, and \eqref{eq:prob1-fin} gives the constraints on the power allocation matrix $\mathbf{P}$.
Solving \eqref{eq:prob1-ini}-\eqref{eq:prob1-fin} means finding the susceptance matrices $\mathbf{B}_F$ and $\mathbf{B}_G$ and the power allocation $p_1,\ldots,p_{N_S}$ that maximize the rate $R$.

\section{Solving the Rate Maximization Problem}
\label{sec:solution}

\begin{figure}[t]
\centering
\includegraphics[width=0.48\textwidth]{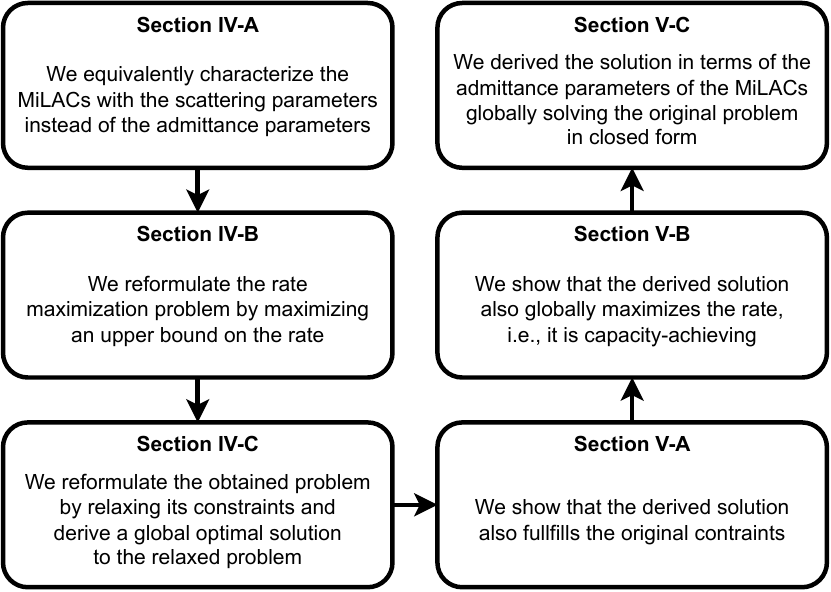}
\caption{Framework for globally solving the rate maximization problem in closed form.}
\label{fig:framework}
\end{figure}

The rate maximization problem in \eqref{eq:prob1-ini}-\eqref{eq:prob1-fin} is inherently non-convex, which complicates its solution.
Due to the non-convex nature of the objective function and the presence of non-convex constraints, standard convex optimization techniques cannot be directly applied.
As a result, finding a global optimal solution is not trivial.
To tackle this problem, in this section, we follow three steps: \textit{i)} we equivalently reformulate the problem by modeling the \glspl{milac} with the scattering parameters, \textit{ii)} we reformulate the problem by maximizing an upper bound on the rate $R$, and \textit{iii)} we reformulate the resulting problem by relaxing its constraints.
Interestingly, we will then prove in Section~\ref{sec:optimal} that the obtained solution is globally optimal for the original problem \eqref{eq:prob1-ini}-\eqref{eq:prob1-fin}, despite being derived by maximizing an upper bound on the rate and by relaxing the constraints.
Fig.~\ref{fig:framework} illustrates the adopted framework that leads to a global optimal closed-form solution to the rate maximization problem.

\subsection{From Admittance to Scattering Parameters Representation}

In Section~\ref{sec:system}, we have characterized the microwave networks of the \glspl{milac} at the transmitter and receiver through their admittance parameters, also known as Y-parameters, i.e., their admittance matrices $\mathbf{Y}_F$ and $\mathbf{Y}_G$, respectively.
This choice has been made since the admittance matrix of a \gls{milac} is closely related to the tunable admittance components in its microwave network.
Nevertheless, according to microwave theory \cite[Chapter~4]{poz11}, there exist multiple equivalent representations that can be used to characterize a microwave network.
For example, in addition to the Y-parameters, a multiport microwave network can also be characterized by the S-parameters, through its scattering matrix.
In detail, a generic $N$-port microwave network with admittance matrix $\mathbf{Y}\in\mathbb{C}^{N\times N}$ can be equivalently characterized by its scattering matrix $\boldsymbol{\Theta}\in\mathbb{C}^{N\times N}$ given by
\begin{equation}
\boldsymbol{\Theta}=\left(Y_0\mathbf{I}_N+\mathbf{Y}\right)^{-1}\left(Y_0\mathbf{I}_N-\mathbf{Y}\right),\label{eq:T(Y)}
\end{equation}
as a function of $\mathbf{Y}$, where $Y_0=Z_0^{-1}$ and $Z_0$ is the reference impedance used to compute the scattering matrix, which is typically set to $Z_0=50~\Omega$ \cite[Chapter~4]{poz11}.
By developing the matrix product in \eqref{eq:T(Y)}, we obtain
\begin{align}
\boldsymbol{\Theta}
=&Y_0\left(Y_0\mathbf{I}_N+\mathbf{Y}\right)^{-1}-\left(Y_0\mathbf{I}_N+\mathbf{Y}\right)^{-1}\mathbf{Y}\\
=&Y_0\left(Y_0\mathbf{I}_N+\mathbf{Y}\right)^{-1}+Y_0\left(Y_0\mathbf{I}_N+\mathbf{Y}\right)^{-1}\\
&-Y_0\left(Y_0\mathbf{I}_N+\mathbf{Y}\right)^{-1}-\left(Y_0\mathbf{I}_N+\mathbf{Y}\right)^{-1}\mathbf{Y}\\
=&2Y_0\left(Y_0\mathbf{I}_N+\mathbf{Y}\right)^{-1}-\mathbf{I}_N,\label{eq:T(B)}
\end{align}
yielding the relationship
\begin{equation}
\left(\frac{\mathbf{Y}}{Y_0}+\mathbf{I}_N\right)^{-1}=\frac{1}{2}\left(\boldsymbol{\Theta}+\mathbf{I}_N\right),\label{eq:alternative}
\end{equation}
which can be used to simplify the expressions of the matrices $\mathbf{F}$ and $\mathbf{G}$ in \eqref{eq:F(Y)2} and \eqref{eq:G(Y)}, respectively.

By introducing the scattering matrix of the \gls{milac} at the transmitter as $\boldsymbol{\Theta}_F\in\mathbb{C}^{(N_S+N_T)\times(N_S+N_T)}$, related to $\mathbf{Y}_F$ via
\begin{equation}
\boldsymbol{\Theta}_F=\left(Y_0\mathbf{I}_{N_S+N_T}+\mathbf{Y}_F\right)^{-1}\left(Y_0\mathbf{I}_{N_S+N_T}-\mathbf{Y}_F\right),\label{eq:TF(YF)}
\end{equation}
because of \eqref{eq:T(Y)}, the precoding matrix $\mathbf{F}$ in \eqref{eq:F(Y)2} can be rewritten as
\begin{equation}
\mathbf{F}=\left[\frac{1}{2}\left(\boldsymbol{\Theta}_F+\mathbf{I}_{N_S+N_T}\right)\right]_{N_S+\mathcal{N}_T,\mathcal{N}_S},
\end{equation}
following the relationship in \eqref{eq:alternative}.
Furthermore, $\mathbf{F}$ can be rewritten as
\begin{equation}
\mathbf{F}=\frac{1}{2}\left[\boldsymbol{\Theta}_F\right]_{N_S+\mathcal{N}_T,\mathcal{N}_S}+\frac{1}{2}\left[\mathbf{I}_{N_S+N_T}\right]_{N_S+\mathcal{N}_T,\mathcal{N}_S},
\end{equation}
and, since $[\mathbf{I}_{N_S+N_T}]_{N_S+\mathcal{N}_T,\mathcal{N}_S}=\mathbf{0}_{N_T\times N_S}$, it simplifies as
\begin{equation}
\mathbf{F}=\frac{1}{2}\left[\boldsymbol{\Theta}_F\right]_{N_S+\mathcal{N}_T,\mathcal{N}_S},
\end{equation}
providing an expression for the precoding matrix simpler than \eqref{eq:F(Y)2}.
A similar discussion can be carried out for the \gls{milac} at the receiver and, by introducing its scattering matrix as $\boldsymbol{\Theta}_G\in\mathbb{C}^{(N_R+N_S)\times(N_R+N_S)}$, which is related to $\mathbf{Y}_G$ via
\begin{equation}
\boldsymbol{\Theta}_G=\left(Y_0\mathbf{I}_{N_R+N_S}+\mathbf{Y}_G\right)^{-1}\left(Y_0\mathbf{I}_{N_R+N_S}-\mathbf{Y}_G\right),\label{eq:TG(YG)}
\end{equation}
the combining matrix $\mathbf{G}$ in \eqref{eq:G(Y)} can be expressed in a simpler way as
\begin{equation}
\mathbf{G}=\frac{1}{2}\left[\boldsymbol{\Theta}_G\right]_{N_R+\mathcal{N}_S,\mathcal{N}_R}.
\end{equation}
In terms of constraints, a lossless microwave network is characterized by a unitary scattering matrix, while a reciprocal microwave network has a symmetric scattering matrix \cite[Chapter~4]{poz11}.
When both these two properties apply, the scattering matrix of a lossless and reciprocal microwave network is unitary and symmetric, as considered to implement \gls{bd-ris} \cite{she22,ner24-1}.
Thus, $\boldsymbol{\Theta}_F$ is mathematically subject to
\begin{equation}
\boldsymbol{\Theta}_F^H\boldsymbol{\Theta}_F=\mathbf{I}_{N_S+N_T},\;\boldsymbol{\Theta}_F=\boldsymbol{\Theta}_F^{T},
\end{equation}
for the \gls{milac} at the transmitter and $\boldsymbol{\Theta}_G$ is subject to
\begin{equation}
\boldsymbol{\Theta}_G^H\boldsymbol{\Theta}_G=\mathbf{I}_{N_R+N_S},\;\boldsymbol{\Theta}_G=\boldsymbol{\Theta}_G^{T},
\end{equation}
for the \gls{milac} at the receiver.

By using the scattering parameters representation, the problem in \eqref{eq:prob1-ini}-\eqref{eq:prob1-fin} can be equivalently rewritten as
\begin{align}
\underset{\boldsymbol{\Theta}_F,\boldsymbol{\Theta}_G,p_{s}}{\mathsf{\mathrm{max}}}\;\;
&R\label{eq:prob2-ini}\\
\mathsf{\mathrm{s.t.}}\;\;\;
&\mathbf{F}=\frac{1}{2}\left[\boldsymbol{\Theta}_F\right]_{N_S+\mathcal{N}_T,\mathcal{N}_S},\\
&\boldsymbol{\Theta}_F^H\boldsymbol{\Theta}_F=\mathbf{I}_{N_S+N_T},\;\boldsymbol{\Theta}_F=\boldsymbol{\Theta}_F^T,\\
&\mathbf{G}=\frac{1}{2}\left[\boldsymbol{\Theta}_G\right]_{N_R+\mathcal{N}_S,\mathcal{N}_R},\\
&\boldsymbol{\Theta}_G^H\boldsymbol{\Theta}_G=\mathbf{I}_{N_R+N_S},\;\boldsymbol{\Theta}_G=\boldsymbol{\Theta}_G^T,\\
&\mathbf{P}=\text{diag}(p_1,\ldots,p_{N_S}),\;\sum_{s=1}^{N_S}p_{s}=1,\label{eq:prob2-fin}
\end{align}
which is solved by determining the scattering matrices $\boldsymbol{\Theta}_F$ and $\boldsymbol{\Theta}_G$ and the power allocation $p_1,\ldots,p_{N_S}$ that maximize the rate $R$.
After finding the optimal $\boldsymbol{\Theta}_F$ and $\boldsymbol{\Theta}_G$, the admittance matrices $\mathbf{Y}_F=j\mathbf{B}_F$ and $\mathbf{Y}_G=j\mathbf{B}_G$ can be readily obtained by inverting the relationships in \eqref{eq:TF(YF)} and \eqref{eq:TG(YG)}, respectively.

\subsection{Maximizing an Upper Bound on the Rate}

To solve problem \eqref{eq:prob2-ini}-\eqref{eq:prob2-fin}, we maximize an upper bound on the rate $R$ by obtaining a global optimal solution.
We will then show in Section~\ref{sec:optimal} that this upper bound coincides with the rate $R$ in the obtained solution, confirming that the rate $R$ is also globally maximized.

The rate $R$ in \eqref{eq:C3} is obtained via a suboptimal detection technique, which treats the residual interference from other streams as noise.
Thus, the rate $R$ is upper bounded by the mutual information between $\mathbf{s}$ and $\mathbf{z}$, denoted as $\mathcal{I}(\mathbf{s},\mathbf{z})$, which can be achieved via a joint detection strategy, i.e., $R\leq\mathcal{I}(\mathbf{s},\mathbf{z})$.
In addition to the inequality $R\leq\mathcal{I}(\mathbf{s},\mathbf{z})$, we have $\mathcal{I}(\mathbf{s},\mathbf{z})\leq\mathcal{I}(\mathbf{s},\mathbf{y})$ because of the data processing inequality, which can be applied since $\mathbf{z}$ defined in \eqref{eq:z} is a function of $\mathbf{y}$ \cite[Chapter~2]{cov99}.
Consequently, we can introduce the mutual information between $\mathbf{s}$ and $\mathbf{y}$ $\mathcal{I}(\mathbf{s},\mathbf{y})$ as an upper bound on the rate $R$, given by
\begin{equation}
\mathcal{I}(\mathbf{s},\mathbf{y})=\log_2\det\left(\mathbf{I}_{N_R}+\frac{P_T}{\sigma^2}\mathbf{H}\mathbf{F}\mathbf{P}\mathbf{F}^H\mathbf{H}^H\right),
\end{equation}
and the problem of maximizing $\mathcal{I}(\mathbf{s},\mathbf{y})$ writes as
\begin{align}
\underset{\boldsymbol{\Theta}_F,p_{s}}{\mathsf{\mathrm{max}}}\;\;
&\log_2\det\left(\mathbf{I}_{N_R}+\frac{P_T}{\sigma^2}\mathbf{H}\mathbf{F}\mathbf{P}\mathbf{F}^H\mathbf{H}^H\right)\label{eq:prob3-ini}\\
\mathsf{\mathrm{s.t.}}\;\;\;
&\mathbf{F}=\frac{1}{2}\left[\boldsymbol{\Theta}_F\right]_{N_S+\mathcal{N}_T,\mathcal{N}_S},\label{eq:prob3-c1}\\
&\boldsymbol{\Theta}_F^H\boldsymbol{\Theta}_F=\mathbf{I}_{N_S+N_T},\;\boldsymbol{\Theta}_F=\boldsymbol{\Theta}_F^T,\label{eq:prob3-c2}\\
&\mathbf{P}=\text{diag}(p_1,\ldots,p_{N_S}),\;\sum_{s=1}^{N_S}p_{s}=1,\label{eq:prob3-fin}
\end{align}
involving only the optimization variables $\boldsymbol{\Theta}_F$ and $p_1,\ldots,p_{N_S}$.

\subsection{Solving a Relaxed Problem}

To solve problem \eqref{eq:prob3-ini}-\eqref{eq:prob3-fin}, we reformulate it by relaxing constraints \eqref{eq:prob3-c1}-\eqref{eq:prob3-c2} and solve the obtained problem through a global optimal solution.
We will then show in Section~\ref{sec:optimal} that the derived global optimal solution of the relaxed problem also fulfills constraints \eqref{eq:prob3-c1}-\eqref{eq:prob3-c2}, confirming that it also globally solves problem \eqref{eq:prob3-ini}-\eqref{eq:prob3-fin}.

Consider the auxiliary variable $\bar{\mathbf{F}}\in\mathbb{C}^{N_T\times N_S}$ defined as $\bar{\mathbf{F}}=[\boldsymbol{\Theta}_F]_{N_S+\mathcal{N}_T,\mathcal{N}_S}$, such that $\mathbf{F}=\bar{\mathbf{F}}/2$.
Since $\bar{\mathbf{F}}$ is a block of a unitary matrix, each of its $N_S$ columns has an $\ell_2$-norm smaller than 1, giving $\Vert\bar{\mathbf{F}}\Vert_F^2\leq N_S$.
Considering this relaxed contraint on $\bar{\mathbf{F}}$, problem \eqref{eq:prob3-ini}-\eqref{eq:prob3-fin} can be reformulated as
\begin{align}
\underset{\bar{\mathbf{F}},p_{s}}{\mathsf{\mathrm{max}}}\;\;
&\log_2\det\left(\mathbf{I}_{N_R}+\frac{P_T}{4\sigma^2}\mathbf{H}\bar{\mathbf{F}}\mathbf{P}\bar{\mathbf{F}}^H\mathbf{H}^H\right)\label{eq:prob4-ini}\\
\mathsf{\mathrm{s.t.}}\;\;\;
&\left\Vert\bar{\mathbf{F}}\right\Vert_F^2\leq N_S,\\
&\mathbf{P}=\text{diag}(p_1,\ldots,p_{N_S}),\;\sum_{s=1}^{N_S}p_{s}=1.\label{eq:prob4-fin}
\end{align}
To solve problem \eqref{eq:prob4-ini}-\eqref{eq:prob4-fin}, there exists a well-known global optimal solution based on multi-eigenmode transmission along the strongest $N_S$ channel eigenmodes with water-filling power allocation \cite[Chapter~5]{cle13}.
Specifically, consider the \gls{svd} of the channel matrix $\mathbf{H}$ as
\begin{equation}
\mathbf{H}=\mathbf{U}\mathbf{\Sigma}\mathbf{V}^H,
\end{equation}
where $\mathbf{U}\in\mathbb{C}^{N_R\times N_R}$ is a unitary matrix containing the left singular vectors of $\mathbf{H}$ in its columns, $\mathbf{\Sigma}\in\mathbb{R}^{N_R\times N_T}$ is a diagonal matrix containing the singular values of $\mathbf{H}$ in its diagonal, denoted as $\sigma_1,\ldots,\sigma_K$, with $K=\min\{N_T,N_R\}$, which are assumed to be nonnegative and in decreasing order, and $\mathbf{V}\in\mathbb{C}^{N_T\times N_T}$ is a unitary matrix containing the right singular vectors of $\mathbf{H}$ in its columns.
Thus, the optimal $\bar{\mathbf{F}}$ is given by the first $N_S$ columns of $\mathbf{V}$, containing the right singular vectors of $\mathbf{H}$ associated with the strongest $N_S$ singular values.
Denoting as $\bar{\mathbf{V}}\in\mathbb{C}^{N_T\times N_S}$ and $\tilde{\mathbf{V}}\in\mathbb{C}^{N_T\times (N_T-N_S)}$ the matrices containing the first $N_S$ columns of $\mathbf{V}$ and the last $N_T-N_S$ columns of $\mathbf{V}$, respectively, such that $\mathbf{V}=[\bar{\mathbf{V}},\tilde{\mathbf{V}}]$, the optimal $\bar{\mathbf{F}}$ writes as $\bar{\mathbf{F}}^\star=\bar{\mathbf{V}}$.
In addition, the optimal power allocations $p_1^\star,\ldots,p_{N_S}^\star$ are given by the water-filling solution
\begin{equation}
p_s^\star=\max\left\{0,\mu-\frac{4\sigma^2}{P_T\lambda_s}\right\}\label{eq:water-filling}
\end{equation}
for $s=1,\ldots,N_S$, where $\mu$ is chosen so as to satisfy the total power constraint $\sum_{s=1}^{N_S}p_{s}=1$ and $\lambda_s=\sigma_s^2$ is the $s$th eigenvalue of $\mathbf{H}\mathbf{H}^H$, for $s=1,\ldots,N_S$.
By substituting the optimal values $\bar{\mathbf{F}}^\star$ and $p_1^\star,\ldots,p_{N_S}^\star$ into the objective function \eqref{eq:prob4-ini}, we obtain that its maximum value is given by
\begin{equation}
C=\sum_{s=1}^{N_S}\log_2\left(1+\frac{P_Tp_{s}^\star\lambda_{s}}{4\sigma^2}\right).\label{eq:Cstar}
\end{equation}

We recall that the optimal value of the objective function in \eqref{eq:Cstar} is not guaranteed to be a tight upper bound on the rate $R$ for two reasons.
First, \eqref{eq:Cstar} is the maximum value of the mutual information $\mathcal{I}(\mathbf{s},\mathbf{y})$ between $\mathbf{s}$ and $\mathbf{y}$, which is an upper bound on the rate $R$.
Second, \eqref{eq:Cstar} has been obtained by considering relaxed constraints on $\mathbf{F}$, given the complexity of its original constraints.
Thus, it is for the moment unclear whether the rate $R$ can achieve the capacity $C$ in \eqref{eq:Cstar} considering the original constraints of $\mathbf{F}$.
In the following, we first show that the derived optimal solution of $\mathbf{F}$ falls within its original constraints.
We then show that with this solution the value of the rate $R$ coincides with the capacity $C$ in \eqref{eq:Cstar}, confirming the feasibility and the global optimality of our solution.

\section{Showing that the Derived Solution\\is Feasible and Global Optimal}
\label{sec:optimal}

We have solved the rate maximization problem and derived a closed-form expression for the maximum rate that can be achieved.
However, the proposed solution has been derived by solving a problem with relaxed constraints, and is not guaranteed to satisfy the constraints of the original optimization problem.
In addition, the proposed solution has been derived by maximizing an upper bound on the rate, and is not guaranteed to globally maximize the rate.
In this section, we address these open questions by proving that the derived solution is feasible, i.e., satisfies the constraints of the original rate maximization problem, and globally maximizes the rate.

\subsection{Checking Solution Feasibility}

We now verify that the solution for the auxiliary variable $\bar{\mathbf{F}}$ given by $\bar{\mathbf{F}}^\star=\bar{\mathbf{V}}$ is compatible with the original constraints \eqref{eq:prob3-c1}-\eqref{eq:prob3-c2}, which require $\bar{\mathbf{F}}$ to be such that $\bar{\mathbf{F}}=[\boldsymbol{\Theta}_F]_{N_R+\mathcal{N}_S,\mathcal{N}_R}$, with $\boldsymbol{\Theta}_F^H\boldsymbol{\Theta}_F=\mathbf{I}_{N_S+N_T}$ and $\boldsymbol{\Theta}_F=\boldsymbol{\Theta}_F^T$.
To this end, we need to find a symmetric unitary matrix $\boldsymbol{\Theta}_F$ having the block $[\boldsymbol{\Theta}_F]_{N_R+\mathcal{N}_S,\mathcal{N}_R}=\bar{\mathbf{V}}$, i.e., a symmetric unitary matrix $\boldsymbol{\Theta}_F$ which can be partitioned as
\begin{equation}
\boldsymbol{\Theta}_F=
\begin{bmatrix}
\mathbf{X}_1 & \bar{\mathbf{V}}^T\\
\bar{\mathbf{V}} & \mathbf{X}_2
\end{bmatrix},\label{eq:TF-X1X2}
\end{equation}
where $\mathbf{X}_1\in\mathbb{C}^{N_S\times N_S}$ and $\mathbf{X}_2\in\mathbb{C}^{N_T\times N_T}$.
This problem can be formalized in the following feasibility-check problem
\begin{align}
\mathrm{find}\;\;
&\mathbf{X}_1\in\mathbb{C}^{N_S\times N_S},\mathbf{X}_2\in\mathbb{C}^{N_T\times N_T}\label{eq:prob5-ini}\\
\mathsf{\mathrm{s.t.}}\;\;\;
&\begin{bmatrix}
\mathbf{X}_1 & \bar{\mathbf{V}}^T\\
\bar{\mathbf{V}} & \mathbf{X}_2
\end{bmatrix}^H
\begin{bmatrix}
\mathbf{X}_1 & \bar{\mathbf{V}}^T\\
\bar{\mathbf{V}} & \mathbf{X}_2
\end{bmatrix}=\mathbf{I}_{N_S+N_T},\\
&\begin{bmatrix}
\mathbf{X}_1 & \bar{\mathbf{V}}^T\\
\bar{\mathbf{V}} & \mathbf{X}_2
\end{bmatrix}=
\begin{bmatrix}
\mathbf{X}_1 & \bar{\mathbf{V}}^T\\
\bar{\mathbf{V}} & \mathbf{X}_2
\end{bmatrix}^T.\label{eq:prob5-fin}
\end{align}
Note that problems involving the completion of symmetric unitary matrices are not trivial due to the joint presence of the two constraints, symmetric and unitary.
For example, a problem similar to \eqref{eq:prob5-ini}-\eqref{eq:prob5-fin} was considered in \cite{mat69} and solved through an iterative solution.
In the following, we show how problem \eqref{eq:prob5-ini}-\eqref{eq:prob5-fin} can be solved in closed form.

We first notice that all the columns of a unitary matrix have unit $\ell_2$-norm and that all the columns of $\bar{\mathbf{V}}$ have also unit norm.
Thus, we must have $\mathbf{X}_1=\mathbf{0}_{N_S}$ to ensure that the first $N_S$ columns of $\boldsymbol{\Theta}_F$ have unit $\ell_2$-norm, and $\mathbf{X}_2$ can be found by solving
\begin{align}
\mathrm{find}\;\;
&\mathbf{X}_2\in\mathbb{C}^{N_S\times N_S}\label{eq:prob6-ini}\\
\mathsf{\mathrm{s.t.}}\;\;\;
&\begin{bmatrix}
\mathbf{0}_{N_S} & \bar{\mathbf{V}}^T\\
\bar{\mathbf{V}} & \mathbf{X}_2
\end{bmatrix}^H
\begin{bmatrix}
\mathbf{0}_{N_S} & \bar{\mathbf{V}}^T\\
\bar{\mathbf{V}} & \mathbf{X}_2
\end{bmatrix}=\mathbf{I}_{N_S+N_T},\label{eq:prob6-c1}\\
&\begin{bmatrix}
\mathbf{0}_{N_S} & \bar{\mathbf{V}}^T\\
\bar{\mathbf{V}} & \mathbf{X}_2
\end{bmatrix}=
\begin{bmatrix}
\mathbf{0}_{N_S} & \bar{\mathbf{V}}^T\\
\bar{\mathbf{V}} & \mathbf{X}_2
\end{bmatrix}^T.\label{eq:prob6-fin}
\end{align}
To solve problem \eqref{eq:prob6-ini}-\eqref{eq:prob6-fin}, we notice that constraint \eqref{eq:prob6-c1} can be equivalently rewritten as
\begin{gather}
\bar{\mathbf{V}}^H\bar{\mathbf{V}}=\mathbf{I}_{N_S},\label{eq:prob6-c1-1}\\
\bar{\mathbf{V}}^H\mathbf{X}_2=\mathbf{0}_{N_S\times N_T},\label{eq:prob6-c1-2}\\
\mathbf{X}_2^H\bar{\mathbf{V}}=\mathbf{0}_{N_T\times N_S},\label{eq:prob6-c1-3}\\
\bar{\mathbf{V}}^*\bar{\mathbf{V}}^T+\mathbf{X}_2^H\mathbf{X}_2=\mathbf{I}_{N_T},\label{eq:prob6-c1-4}
\end{gather}
where \eqref{eq:prob6-c1-1} is always satisfied and \eqref{eq:prob6-c1-3} coincides with \eqref{eq:prob6-c1-2} since they are the Hermitian of each other.
Besides, constraint \eqref{eq:prob6-fin} is satisfied if and only if
\begin{equation}
\mathbf{X}_2=\mathbf{X}_2^T.\label{eq:prob6-c2-1}
\end{equation}
Thus, problem \eqref{eq:prob6-ini}-\eqref{eq:prob6-fin} can be equivalently reformulated as
\begin{align}
\mathrm{find}\;\;
&\mathbf{X}_2\in\mathbb{C}^{N_S\times N_S}\label{eq:prob7-ini}\\
\mathsf{\mathrm{s.t.}}\;\;\;
&\eqref{eq:prob6-c1-2},\;\eqref{eq:prob6-c1-4},\;\eqref{eq:prob6-c2-1},\label{eq:prob7-fin}
\end{align}
with a solution given by $\mathbf{X}_2=-\tilde{\mathbf{V}}\tilde{\mathbf{V}}^T$, as it can be verified by showing that all three constraints are fulfilled by this solution.
First, $\bar{\mathbf{V}}^H\mathbf{X}_2=-\bar{\mathbf{V}}^H\tilde{\mathbf{V}}\tilde{\mathbf{V}}^T=\mathbf{0}_{N_S\times N_T}$ since all the columns of $\tilde{\mathbf{V}}$ are orthogonal to all the columns of $\bar{\mathbf{V}}$, which gives $\bar{\mathbf{V}}^H\tilde{\mathbf{V}}=\mathbf{0}_{N_S\times N_T}$.
Second, $\bar{\mathbf{V}}^*\bar{\mathbf{V}}^T+\mathbf{X}_2^H\mathbf{X}_2=\bar{\mathbf{V}}^*\bar{\mathbf{V}}^T+\tilde{\mathbf{V}}^*\tilde{\mathbf{V}}^T=\mathbf{V}^*\mathbf{V}^T=(\mathbf{V}^H\mathbf{V})^*$, yielding $\bar{\mathbf{V}}^*\bar{\mathbf{V}}^T+\mathbf{X}_2^H\mathbf{X}_2=\mathbf{I}_{N_T}$ since $\mathbf{V}$ is unitary and $(\mathbf{V}^H\mathbf{V})^*=\mathbf{I}_{N_T}$.
Third, it is obvious that $-\tilde{\mathbf{V}}\tilde{\mathbf{V}}^T$ is symmetric.
Note that $\mathbf{X}_2=-\tilde{\mathbf{V}}\tilde{\mathbf{V}}^T$ is just a valid solution among infinitely many others\footnote{While also $\mathbf{X}_2=\tilde{\mathbf{V}}\tilde{\mathbf{V}}^T$ solves problem \eqref{eq:prob7-ini}-\eqref{eq:prob7-fin}, we adopt $\mathbf{X}_2=-\tilde{\mathbf{V}}\tilde{\mathbf{V}}^T$ since it allows to express the admittance matrix of the \gls{milac} at the transmitter in closed form, as it will be clarified in Section~\ref{sec:StoY}.}.
For example, any matrix $\mathbf{X}_2=\mathbf{A}\mathbf{A}^T$, with $\mathbf{A}\in\mathbb{C}^{N_T\times (N_T-N_S)}$, is a valid solution if all the columns of $\mathbf{A}$ have unit $\ell_2$-norm, are orthogonal between themselves and are orthogonal with the $N_S$ columns of $\bar{\mathbf{V}}$.

Considering the derived solution to problem \eqref{eq:prob5-ini}-\eqref{eq:prob5-fin}, a symmetric unitary scattering matrix $\boldsymbol{\Theta}_F$ having $[\boldsymbol{\Theta}_F]_{N_R+\mathcal{N}_S,\mathcal{N}_R}=\bar{\mathbf{V}}$ can be obtained by substituting $\mathbf{X}_1=\mathbf{0}_{N_S}$ and $\mathbf{X}_2=-\tilde{\mathbf{V}}\tilde{\mathbf{V}}^T$ in \eqref{eq:TF-X1X2} as
\begin{equation}
\boldsymbol{\Theta}_F^\star=
\begin{bmatrix}
\mathbf{0}_{N_S} & \bar{\mathbf{V}}^T\\
\bar{\mathbf{V}} & -\tilde{\mathbf{V}}\tilde{\mathbf{V}}^T
\end{bmatrix}.\label{eq:TF}
\end{equation}
This proves that the global optimal solution to the relaxed problem \eqref{eq:prob4-ini}-\eqref{eq:prob4-fin} is also a valid global optimal solution to the problem \eqref{eq:prob3-ini}-\eqref{eq:prob3-fin}.

\subsection{Checking Global Maximization of the Rate}

We have derived a global optimal solution to the problem \eqref{eq:prob3-ini}-\eqref{eq:prob3-fin}.
However, the objective function in \eqref{eq:prob3-ini} is an upper bound on the rate $R$, and there is no guarantee that the obtained solution also globally maximizes the rate $R$.
In addition, we only have a solution for the scattering matrix of the \gls{milac} at the transmitter $\boldsymbol{\Theta}_F$ and the power allocations $p_s$, for $s=1,\ldots,N_S$, while it is still unclear how to reconfigure the scattering matrix of the \gls{milac} at the receiver $\boldsymbol{\Theta}_G$.
We now propose a solution for $\boldsymbol{\Theta}_G$ and verify that, jointly with the previously derived solutions for $\boldsymbol{\Theta}_F$ and $p_s$, they indeed globally maximize the rate $R$.

Adopting a dual approach to that employed at the transmitter side, we reconfigure the \gls{milac} at the receiver such that the combining matrix $\mathbf{G}$ contains in its $N_S$ rows the left singular vectors of $\mathbf{H}$ associated with the strongest $N_S$ singular values.
More formally, denoting as $\bar{\mathbf{U}}\in\mathbb{C}^{N_R\times N_S}$ and $\tilde{\mathbf{U}}\in\mathbb{C}^{N_R\times (N_R-N_S)}$ the matrices containing the first $N_S$ columns of $\mathbf{U}$ and the last $N_R-N_S$ columns of $\mathbf{U}$, respectively, such that $\mathbf{U}=[\bar{\mathbf{U}},\tilde{\mathbf{U}}]$, we speculate that the optimal $\mathbf{G}$ maximizing the rate $R$ is given by $\mathbf{G}^\star=\bar{\mathbf{U}}^H/2$.
Thus, a possible scattering matrix of the \gls{milac} at the receiver $\boldsymbol{\Theta}_G$ satisfying $[\boldsymbol{\Theta}_G]_{N_R+\mathcal{N}_S,\mathcal{N}_R}=\bar{\mathbf{U}}^H$ and fulfilling the symmetric unitary constraints is
\begin{equation}
\boldsymbol{\Theta}_G^\star=
\begin{bmatrix}
-\tilde{\mathbf{U}}^*\tilde{\mathbf{U}}^H & \bar{\mathbf{U}}^*\\
\bar{\mathbf{U}}^H & \mathbf{0}_{N_S} 
\end{bmatrix},\label{eq:TG}
\end{equation}
as it can be easily verified by checking that $\boldsymbol{\Theta}_G^{\star H}\boldsymbol{\Theta}_G^\star=\mathbf{I}_{N_R+N_S}$ and $\boldsymbol{\Theta}_G^\star=\boldsymbol{\Theta}_G^{\star T}$.

With the solutions of the scattering matrices $\boldsymbol{\Theta}_F^\star$ in \eqref{eq:TF} and $\boldsymbol{\Theta}_G^\star$ in \eqref{eq:TG}, the optimal precoding and combining matrices are given by $\mathbf{F}^\star=\bar{\mathbf{V}}/2$ and $\mathbf{G}^\star=\bar{\mathbf{U}}^H/2$, respectively.
This gives $\mathbf{G}^\star\mathbf{H}\mathbf{F}^\star=[\mathbf{\Sigma}]_{\mathcal{N}_S,\mathcal{N}_S}/4$, which simplifies the rate in \eqref{eq:C1} to
\begin{equation}
R=\sum_{s=1}^{N_S}\log_2\left(1+\frac{P_Tp_{s}\lambda_s}
{4\sigma^2}\right),\label{eq:Copt}
\end{equation}
since $\vert[\mathbf{G}^\star\mathbf{H}\mathbf{F}^\star]_{s,s}\vert^2=\lambda_s/16$, $\vert[\mathbf{G}^\star\mathbf{H}\mathbf{F}^\star]_{s,t}\vert^2=0$ if $t\neq s$, and $\Vert[\mathbf{G}^\star]_{s,:}\Vert^2=1/4$, for $s,t=1,\ldots,N_S$.
By also considering the optimal power allocation $p_1^\star,\ldots,p_{N_S}^\star$ given by the water-filling solution in \eqref{eq:water-filling}, the achieved rate in \eqref{eq:Copt} coincides with the capacity in \eqref{eq:Cstar}, proving that the obtained solution is globally optimal.

\subsection{From Scattering to Admittance Parameters Representation}
\label{sec:StoY}

We have determined the expressions for the optimal scattering matrices of the \glspl{milac} at the transmitter and receiver, in \eqref{eq:TF} and \eqref{eq:TG}, respectively, which are proved to globally solve problem \eqref{eq:prob2-ini}-\eqref{eq:prob2-fin}.
However, the original problem \eqref{eq:prob1-ini}-\eqref{eq:prob1-fin} involves the optimization of the admittance matrices of the \glspl{milac}, which are directly related to their tunable admittance components.
In the following, we derive the expressions of the optimal admittance matrices of the \glspl{milac}, concluding the solution of problem \eqref{eq:prob1-ini}-\eqref{eq:prob1-fin}.

For the \gls{milac} at the transmitter, its optimal susceptance matrix $\mathbf{B}_F^\star$ is related to its optimal scattering matrix $\boldsymbol{\Theta}_F^\star$ through
\begin{equation}
\mathbf{B}_F^\star=-jY_0\left(2\left(\boldsymbol{\Theta}_F^\star+\mathbf{I}_{N_S+N_T}\right)^{-1}-\mathbf{I}_{N_S+N_T}\right),\label{eq:BF}
\end{equation}
which is a direct consequence of \eqref{eq:TF(YF)} and that $\mathbf{Y}_F=j\mathbf{B}_F$ in a lossless \gls{milac}.
By substituting $\boldsymbol{\Theta}_F^\star$ given by \eqref{eq:TF} into \eqref{eq:BF}, we obtain
\begin{equation}
\mathbf{B}_F^\star=-jY_0\left(2
\begin{bmatrix}
\mathbf{I}_{N_S} & \bar{\mathbf{V}}^T\\
\bar{\mathbf{V}} & \mathbf{I}_{N_T}-\tilde{\mathbf{V}}\tilde{\mathbf{V}}^T
\end{bmatrix}^{-1}-\mathbf{I}_{N_S+N_T}\right),\label{eq:BF-inv}
\end{equation}
which can also be written as
\begin{equation}
\mathbf{B}_F^\star=-jY_0\left(2
\begin{bmatrix}
\mathbf{K}_{F,11} & \mathbf{K}_{F,12}\\
\mathbf{K}_{F,21} & \mathbf{K}_{F,22}
\end{bmatrix}-\mathbf{I}_{N_S+N_T}\right),\label{eq:BF-K}
\end{equation}
with
\begin{gather}
\mathbf{K}_{F,11}=\mathbf{I}_{N_S}+\bar{\mathbf{V}}^T\left(\mathbf{I}_{N_T}-\mathbf{V}\mathbf{V}^T\right)^{-1}\bar{\mathbf{V}},\label{eq:KF11}\\
\mathbf{K}_{F,12}=-\bar{\mathbf{V}}^T\left(\mathbf{I}_{N_T}-\mathbf{V}\mathbf{V}^T\right)^{-1},\\
\mathbf{K}_{F,21}=-\left(\mathbf{I}_{N_T}-\mathbf{V}\mathbf{V}^T\right)^{-1}\bar{\mathbf{V}},\\
\mathbf{K}_{F,22}=\left(\mathbf{I}_{N_T}-\mathbf{V}\mathbf{V}^T\right)^{-1},\label{eq:KF22}
\end{gather}
where we exploited the expression of the inverse of a $2\times2$ block matrix \cite{lu02} and the relationship $\bar{\mathbf{V}}\bar{\mathbf{V}}^T+\tilde{\mathbf{V}}\tilde{\mathbf{V}}^T=\mathbf{V}\mathbf{V}^T$, valid for any matrix $\mathbf{V}=[\bar{\mathbf{V}},\tilde{\mathbf{V}}]$.
After the necessary computations, the expressions in \eqref{eq:KF11}-\eqref{eq:KF22} can be rewritten as functions of only $\Re\{\mathbf{V}\}$ and $\Im\{\mathbf{V}\}$ as
\begin{gather}
\begin{split}
\mathbf{I}_{N_S}+\bar{\mathbf{V}}^T&\left(\mathbf{I}_{N_T}-\mathbf{V}\mathbf{V}^T\right)^{-1}\bar{\mathbf{V}}\\
&=\frac{1}{2}\left(\mathbf{I}_{N_S}+j\left[\Im\{\mathbf{V}\}^{-1}\Re\{\mathbf{V}\}\right]_{\mathcal{N}_S,\mathcal{N}_S}\right),
\end{split}\label{eq:KF11-sim}\\
-\bar{\mathbf{V}}^T\left(\mathbf{I}_{N_T}-\mathbf{V}\mathbf{V}^T\right)^{-1}
=-\frac{1}{2}j\left[\Im\{\mathbf{V}\}^{-1}\right]_{\mathcal{N}_S,:},\\
-\left(\mathbf{I}_{N_T}-\mathbf{V}\mathbf{V}^T\right)^{-1}\bar{\mathbf{V}}
=-\frac{1}{2}j\left[\Im\{\mathbf{V}\}^{-1}\right]_{\mathcal{N}_S,:}^T,\\
\left(\mathbf{I}_{N_T}-\mathbf{V}\mathbf{V}^T\right)^{-1}
=\frac{1}{2}\left(\mathbf{I}_{N_T}+j\Re\{\mathbf{V}\}\Im\{\mathbf{V}\}^{-1}\right),\label{eq:KF22-sim}
\end{gather}
respectively.
Thus, by substituting \eqref{eq:KF11-sim}-\eqref{eq:KF22-sim} into \eqref{eq:KF11}-\eqref{eq:KF22}, and then into \eqref{eq:BF-K}, we obtain $\mathbf{B}_F^\star$ in closed-form as
\begin{equation}
\mathbf{B}_F^\star=Y_0
\begin{bmatrix}
\left[\Im\{\mathbf{V}\}^{-1}\Re\{\mathbf{V}\}\right]_{\mathcal{N}_S,\mathcal{N}_S} &
-\left[\Im\{\mathbf{V}\}^{-1}\right]_{\mathcal{N}_S,:} \\
-\left[\Im\{\mathbf{V}\}^{-1}\right]_{\mathcal{N}_S,:}^T &
\Re\{\mathbf{V}\}\Im\{\mathbf{V}\}^{-1}
\end{bmatrix},\label{eq:BFstar}
\end{equation}
purely as a function of $\Re\{\mathbf{U}\}$ and $\Im\{\mathbf{V}\}$.
As expected for the susceptance matrix of a reciprocal microwave network, note that $\mathbf{B}_F^\star$ is symmetric since the matrices $\Im\{\mathbf{V}\}^{-1}\Re\{\mathbf{V}\}$ and $\Re\{\mathbf{V}\}\Im\{\mathbf{V}\}^{-1}$ are symmetric for any unitary matrix $\mathbf{V}$ with invertible imaginary part.

A similar discussion can be repeated for the \gls{milac} at the receiver by departing from the relationship  between its optimal admittance matrix $\mathbf{B}_G^\star$ and its optimal scattering matrix $\boldsymbol{\Theta}_G^\star$, given by
\begin{equation}
\mathbf{B}_G^\star=-jY_0\left(2\left(\boldsymbol{\Theta}_G^\star+\mathbf{I}_{N_R+N_S}\right)^{-1}-\mathbf{I}_{N_R+N_S}\right),\label{eq:BG}
\end{equation}
following from \eqref{eq:TG(YG)}, which becomes
\begin{equation}
\mathbf{B}_G^\star=-jY_0\left(2
\begin{bmatrix}
\mathbf{I}_{N_R}-\tilde{\mathbf{U}}^*\tilde{\mathbf{U}}^H & \bar{\mathbf{U}}^*\\
\bar{\mathbf{U}}^H & \mathbf{I}_{N_S}
\end{bmatrix}^{-1}-\mathbf{I}_{N_R+N_S}\right),\label{eq:BG-inv}
\end{equation}
by considering the expression of $\boldsymbol{\Theta}_G^\star$ given by \eqref{eq:TG}.
In addition, by recalling the expression of the inverse of a $2\times2$ block matrix \cite{lu02}, \eqref{eq:BG-inv} can be written as
\begin{equation}
\mathbf{B}_G^\star=-jY_0\left(2
\begin{bmatrix}
\mathbf{K}_{G,11} & \mathbf{K}_{G,12}\\
\mathbf{K}_{G,21} & \mathbf{K}_{G,22}
\end{bmatrix}-\mathbf{I}_{N_R+N_S}\right),\label{eq:BG-K}
\end{equation}
with
\begin{gather}
\mathbf{K}_{G,11}=\left(\mathbf{I}_{N_R}-\mathbf{U}^*\mathbf{U}^H\right)^{-1},\label{eq:KG11}\\
\mathbf{K}_{G,12}=-\left(\mathbf{I}_{N_R}-\mathbf{U}^*\mathbf{U}^H\right)^{-1}\bar{\mathbf{U}}^*,\\
\mathbf{K}_{G,21}=-\bar{\mathbf{U}}^H\left(\mathbf{I}_{N_R}-\mathbf{U}^*\mathbf{U}^H\right)^{-1},\\
\mathbf{K}_{G,22}=\mathbf{I}_{N_S}+\bar{\mathbf{U}}^H\left(\mathbf{I}_{N_R}-\mathbf{U}^*\mathbf{U}^H\right)^{-1}\bar{\mathbf{U}}^*,\label{eq:KG22}
\end{gather}
where we also used the relationship $\bar{\mathbf{U}}^*\bar{\mathbf{U}}^H+\tilde{\mathbf{U}}^*\tilde{\mathbf{U}}^H=\mathbf{U}^*\mathbf{U}^H$, valid for any matrix $\mathbf{U}=[\bar{\mathbf{U}},\tilde{\mathbf{U}}]$.
Following appropriate calculations, \eqref{eq:KG11}-\eqref{eq:KG22} can be expressed as
\begin{gather}
\left(\mathbf{I}_{N_R}-\mathbf{U}^*\mathbf{U}^H\right)^{-1}
=\frac{1}{2}\left(\mathbf{I}_{N_R}-j\Re\{\mathbf{U}\}\Im\{\mathbf{U}\}^{-1}\right),\label{eq:KG11-sim}\\
-\left(\mathbf{I}_{N_R}-\mathbf{U}^*\mathbf{U}^H\right)^{-1}\bar{\mathbf{U}}^*
=\frac{1}{2}j\left[\Im\{\mathbf{U}\}^{-1}\right]_{\mathcal{N}_S,:}^T,\\
-\bar{\mathbf{U}}^H\left(\mathbf{I}_{N_R}-\mathbf{U}^*\mathbf{U}^H\right)^{-1}
=\frac{1}{2}j\left[\Im\{\mathbf{U}\}^{-1}\right]_{\mathcal{N}_S,:},\label{eq:KG22-sim}\\
\begin{split}
\mathbf{I}_{N_S}+\bar{\mathbf{U}}^H&\left(\mathbf{I}_{N_R}-\mathbf{U}^*\mathbf{U}^H\right)^{-1}\bar{\mathbf{U}}^*\\
&=\frac{1}{2}\left(\mathbf{I}_{N_S}-j\left[\Im\{\mathbf{U}\}^{-1}\Re\{\mathbf{U}\}\right]_{\mathcal{N}_S,\mathcal{N}_S}\right),
\end{split}
\end{gather}
allowing us to rewrite \eqref{eq:BG-K} in closed-form as
\begin{equation}
\mathbf{B}_G^\star=Y_0
\begin{bmatrix}
-\Re\{\mathbf{U}\}\Im\{\mathbf{U}\}^{-1} &
\left[\Im\{\mathbf{U}\}^{-1}\right]_{\mathcal{N}_S,:}^T \\
\left[\Im\{\mathbf{U}\}^{-1}\right]_{\mathcal{N}_S,:} &
-\left[\Im\{\mathbf{U}\}^{-1}\Re\{\mathbf{U}\}\right]_{\mathcal{N}_S,\mathcal{N}_S}
\end{bmatrix},\label{eq:BGstar}
\end{equation}
as a function of only $\Re\{\mathbf{U}\}$ and $\Im\{\mathbf{U}\}$.

In summary, the expressions of the optimal susceptance matrices $\mathbf{B}_F^\star$ in \eqref{eq:BFstar} and $\mathbf{B}_G^\star$ in \eqref{eq:BGstar}, together with the optimal power allocation given by the water-filling solution in \eqref{eq:water-filling}, globally solve in closed form the rate maximization problem formulated in \eqref{eq:prob1-ini}-\eqref{eq:prob1-fin}.
This solution allows us to achieve the capacity given in \eqref{eq:Cstar}.

\section{Comparison Between MiLAC-Aided Beamforming and Digital Beamforming}
\label{sec:digital}

\begin{figure}[t]
\centering
\includegraphics[height=0.268\textwidth]{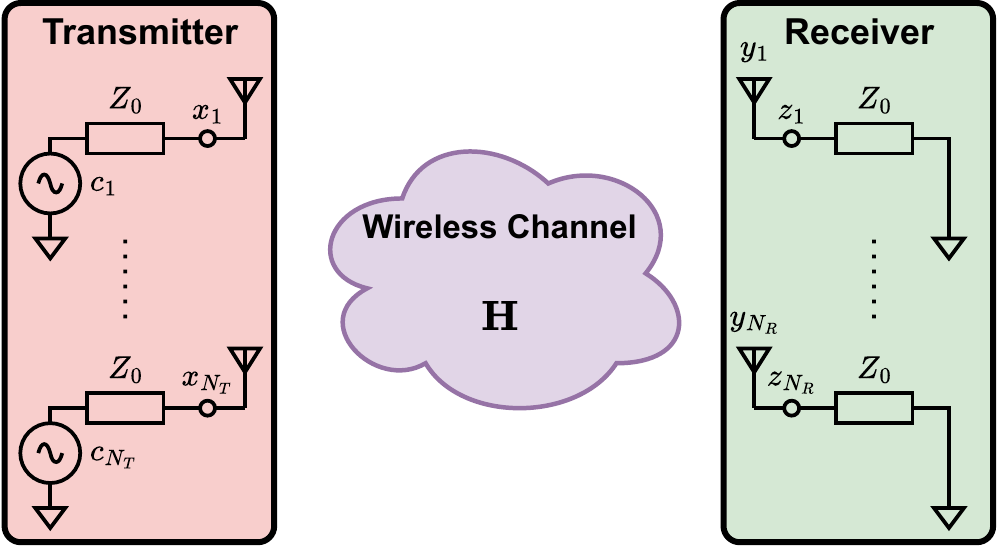}
\caption{Digital MIMO system.}
\label{fig:digital}
\end{figure}

We have solved the rate maximization problem for a \gls{milac}-aided \gls{mimo} system by deriving a global optimal solution in closed form.
We have also characterized the capacity that can be achieved with \gls{milac}-aided beamforming.
In this section, we relate the capacity achievable with \gls{milac}-aided beamforming to that achievable with digital beamforming.

Consider a point-to-point \gls{mimo} system between an $N_T$-antenna transmitter and an $N_R$-antenna receiver as represented in Fig.~\ref{fig:digital}, where both transmitter and receiver operate fully digital beamforming.
To enable fully digital beamforming, the transmitter is equipped with $N_T$ \gls{rf} chains, each connected to a transmitting antenna, and the receiver is equipped with $N_R$ \gls{rf} chains, each connected to a receiving antenna.
We assume that the transmitter and receiver communicate by sending $N_S$ symbols in parallel, i.e., $N_S$ streams, where $N_S\leq\min\{N_T,N_R\}$.
Note that, while the number of streams $N_S$ should be optimized to achieve the information-theoretic capacity \cite{tel99}, it is fixed here to ensure a fair comparison with the \gls{milac}-aided \gls{mimo} system.

The source signal $\mathbf{c}\in\mathbb{C}^{N_T\times1}$ at the voltage generators contains the precoded $N_S$ transmitted symbols, i.e.,
\begin{equation}
\mathbf{c}=\sqrt{P_T}\mathbf{W}\mathbf{s},\label{eq:c-digital}
\end{equation}
where $\mathbf{s}\in\mathbb{C}^{N_S\times1}$ is the symbol vector such that $\mathbb{E}[\mathbf{s}\mathbf{s}^H]=\mathbf{I}_{N_S}$, $\mathbf{W}\in\mathbb{C}^{N_T\times N_S}$ is the precoding matrix subject to the power constraint $\Vert\mathbf{W}\Vert_F^2=1$, and $P_T$ is the transmitted power.
With these constraints, the average power of the signal at the voltage generators $\mathbf{c}$ is $P_T$, as it can be verified by computing $\mathbb{E}[\Vert\mathbf{c}\Vert^2]=\mathbb{E}[\Vert\sqrt{P_T}\mathbf{W}\mathbf{s}\Vert^2]=P_T\mathbb{E}[\text{Tr}(\mathbf{W}\mathbf{s}\mathbf{s}^H\mathbf{W}^H)]=P_T\text{Tr}(\mathbf{W}\mathbf{W}^H)=P_T\Vert\mathbf{W}\Vert_F^2=P_T$.
Note that this is the same power constraint imposed on the signal at the voltage generators $\mathbf{c}$ in the \gls{milac}-aided \gls{mimo} system introduced in Section~\ref{sec:system}.
While $\mathbf{c}$ is the signal of the source generators, it differs from the signal at the transmit antennas.
In fact, the signal at the transmit antennas $\mathbf{x}\in\mathbb{C}^{N_T\times1}$ is given by $\mathbf{x}=\mathbf{c}/2$ in the case the antennas are perfectly matched to the source generators series impedance $Z_0$, as clarified in the Appendix.

The signal at the receiving antennas $\mathbf{y}\in\mathbb{C}^{N_R\times1}$ writes as
\begin{equation}
\mathbf{y}=\mathbf{H}\mathbf{x}+\mathbf{n},\label{eq:y-digital}
\end{equation}
where $\mathbf{H}\in\mathbb{C}^{N_R\times N_T}$ is the wireless channel and $\mathbf{n}\in\mathbb{C}^{N_R\times 1}$ is the \gls{awgn} such that $\mathbb{E}[\mathbf{n}\mathbf{n}^H]=\sigma^2\mathbf{I}_{N_R}$, with $\sigma^2$ denoting the noise power.
At the receiver side, the signal at the receiving antennas is different from the signal read at the receiving \gls{rf} chains.
Specifically, the signal at the receiving \gls{rf} chains is given by $\mathbf{z}=\mathbf{y}/2$ when the receiving antennas are perfectly matched to $Z_0$, as clarified in the Appendix.
Thus, the end-to-end system model is given by
\begin{equation}
\mathbf{z}=\frac{\sqrt{P_T}}{4}\mathbf{H}\mathbf{W}\mathbf{s}+\frac{1}{2}\mathbf{n}.\label{eq:system-model-digital}
\end{equation}

In this fully digital system, the signal $\mathbf{z}$ can be properly combined at the receiver to decode the transmitted symbols $\mathbf{s}$.
Thus, the rate is given by the mutual information between $\mathbf{s}$ and $\mathbf{z}$, i.e.,
\begin{equation}
R=\log_2\det\left(\mathbf{I}_{N_R}+\frac{P_T}{4\sigma^2}\mathbf{H}\mathbf{W}\mathbf{W}^H\mathbf{H}^H\right),\label{eq:capacity-digital}
\end{equation}
and the rate maximization problem reads as
\begin{align}
\underset{\mathbf{W}}{\mathsf{\mathrm{max}}}\;\;
&R\label{eq:prob8-ini}\\
\mathsf{\mathrm{s.t.}}\;\;\;
&\left\Vert\mathbf{W}\right\Vert_F^2=1,\label{eq:prob8-fin}
\end{align}
in which the precoding matrix $\mathbf{W}$ is optimized to maximize the rate.
It is well-known from \gls{mimo} theory that a global optimal solution to \eqref{eq:prob8-ini}-\eqref{eq:prob8-fin} is given by $\mathbf{W}^\star=\bar{\mathbf{V}}\mathbf{P}^{\star1/2}$, where $\mathbf{P}^{\star1/2}=\text{diag}(\sqrt{p_1^\star},\ldots,\sqrt{p_{N_S}^\star})$ and $p_s^\star$ is set by the water-filling solution as in \eqref{eq:water-filling}, for $s=1,\ldots,N_S$, where $\mu$ is chosen so as to ensure $\sum_{s=1}^{N_S}p_{s}=1$, which satisfies the constraint $\Vert\mathbf{W}\Vert^2_F=1$ \cite[Chapter~5]{cle13}.
By substituting the solution $\mathbf{W}^\star$ into the rate of the \gls{mimo} system operating digital beamforming in \eqref{eq:capacity-digital}, we obtain that its optimal value is given by the capacity $C$ provided in \eqref{eq:Cstar}.
This shows that \gls{milac}-aided beamforming can achieve the same capacity as digital beamforming with the same number of parallel streams $N_S$.
Note that the expression in \eqref{eq:Cstar} is the capacity of a \gls{mimo} system with a fixed number of streams $N_S$, which corresponds to the information-theoretic definition of capacity only when the channel matrix $\mathbf{H}$ has rank $N_S$ \cite{tel99}.

\section{Numerical Results}
\label{sec:results}

\begin{figure}[t]
\centering
\includegraphics[width=0.44\textwidth]{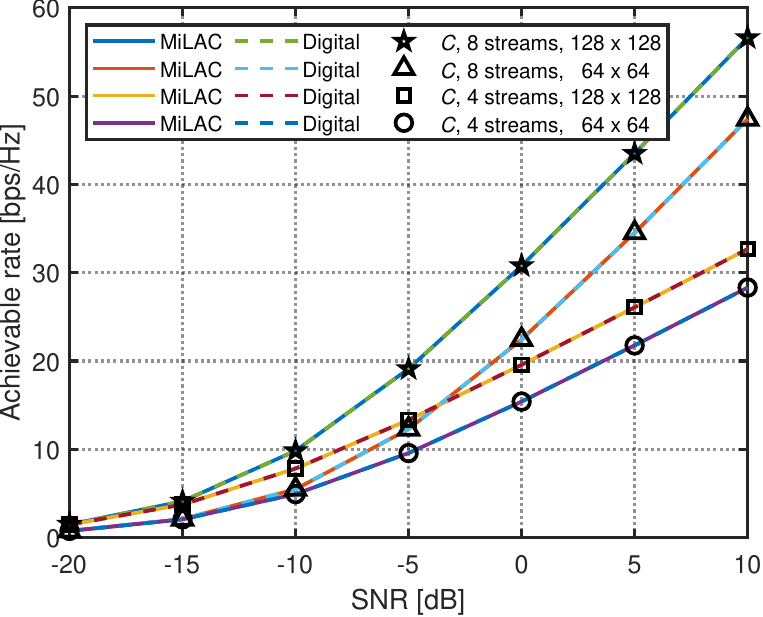}
\caption{Achievable rate versus the SNR in an $N_S$-stream $N_R\times N_T$ MIMO system.
The rate of MiLAC-aided beamforming, digital beamforming, and the capacity in \eqref{eq:Cstar} are compared.}
\label{fig:capacity-snr}
\end{figure}

In this section, we validate the global optimal solution to the rate maximization problem derived in this study, and compare the capacity achieved with \gls{milac}-aided beamforming and digital beamforming.
To this end, we provide numerical results obtained with Monte Carlo simulations.

\begin{figure}[t]
\centering
\includegraphics[width=0.44\textwidth]{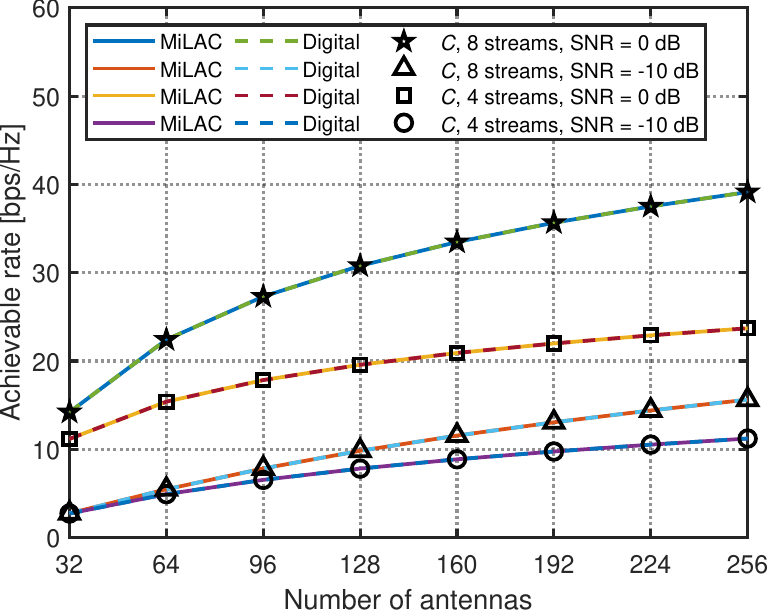}
\caption{Achievable rate versus the number of antennas $N_T=N_R$ in an $N_S$-stream MIMO system with given SNR.
The rate of MiLAC-aided beamforming, digital beamforming, and the capacity in \eqref{eq:Cstar} are compared.}
\label{fig:capacity-antennas}
\end{figure}

In Fig.~\ref{fig:capacity-snr}, we report the achievable rate averaged over multiple channel realizations \gls{iid} Rayleigh distributed versus the \gls{snr} $P_T/\sigma^2$, considering four different scenarios where there are $N_S\in\{4,8\}$ streams and $N_T=N_R\in\{64,128\}$ antennas.
In addition, in Fig.~\ref{fig:capacity-antennas}, we report the average achievable rate with \gls{iid} Rayleigh channels versus the number of antennas at the transmitter and the receiver $N_T=N_R$, considering four different scenarios where there are $N_S\in\{4,8\}$ streams and the \gls{snr} is $P_T/\sigma^2\in\{-10,0\}$~dB.
In Figs.~\ref{fig:capacity-snr} and \ref{fig:capacity-antennas}, for each scenario we compare \textit{i)} ``MiLAC'': the rate achievable with \gls{milac}-aided beamforming using the global optimal solution derived in Sections~\ref{sec:solution} and \ref{sec:optimal}, \textit{ii)} ``Digital'': the rate achievable with digital beamforming as discussed in Section~\ref{sec:digital}, and \textit{iii)} ``$C$'': the capacity given by \eqref{eq:Cstar}.

From Figs.~\ref{fig:capacity-snr} and \ref{fig:capacity-antennas}, we make the following two observations.
\textit{First}, the achievable rate increases with the \gls{snr}, the number of streams $N_S$, and the number of antennas $N_T=N_R$, as well-known from \gls{mimo} theory \cite[Chapter~5]{cle13}.
\textit{Second}, \gls{milac}-aided beamforming exactly achieves the capacity in \eqref{eq:Cstar}, which is the same capacity achieved by digital beamforming, confirming our theoretical insights.
Thus, \gls{milac}-aided beamforming provides maximum performance even under the practical constraints of lossless and reciprocal \glspl{milac}.

\section{Conclusion}
\label{sec:conclusion}

We consider a point-to-point \gls{mimo} system where both transmitter and receiver operate \gls{milac}-aided beamforming with the \glspl{milac} constrained to be lossless and reciprocal for practical reasons.
After characterizing the end-to-end system model and the rate achieved by this \gls{mimo} system fully operating in the analog domain, we set up the rate maximization problem.
In this problem, we aim at reconfiguring the admittance matrices of the \glspl{milac} at the transmitter and receiver to maximize the rate.
Remarkably, the considered optimization problem can be solved in closed form through a solution that is proved to be globally optimal, despite being non-convex.

We also rigorously compare \gls{milac}-aided beamforming with digital beamforming using microwave theory and antenna theory.
As a result of this analysis, we show that the capacity achieved by \gls{milac}-aided beamforming is the same as the capacity achieved by digital beamforming with the same number of streams.
This confirms that \gls{milac}-aided beamforming offers maximum flexibility and performance, even in the practical case of lossless and reciprocal \glspl{milac}.

In conclusion, we can identify the following four benefits offered by \gls{milac}-aided beamforming with lossless and reciprocal \glspl{milac}.
\textit{First}, it achieves maximum capacity, identical to that of digital beamforming with the same number of streams.
\textit{Second}, it minimizes the number of \gls{rf} chains since it only needs as many \gls{rf} chains as the number of streams.
\textit{Third}, it only requires low resolution \glspl{adc}/\glspl{dac} since the symbols are directly carried and detected on the \gls{rf} chains.
\textit{Fourth}, it reduces the computational complexity of beamforming since the symbols are precoded and combined in the analog domain, thereby saving a matrix-vector product per symbol time.

In this work, we have considered a point-to-point \gls{mimo} system and optimized lossless and reciprocal \glspl{milac} to maximize the rate and achieve capacity.
Extending this framework to optimize lossless and reciprocal \glspl{milac} in multi-user systems remains an important direction for future research.

\section*{Appendix}

\begin{figure}[t]
\centering
\includegraphics[width=0.38\textwidth]{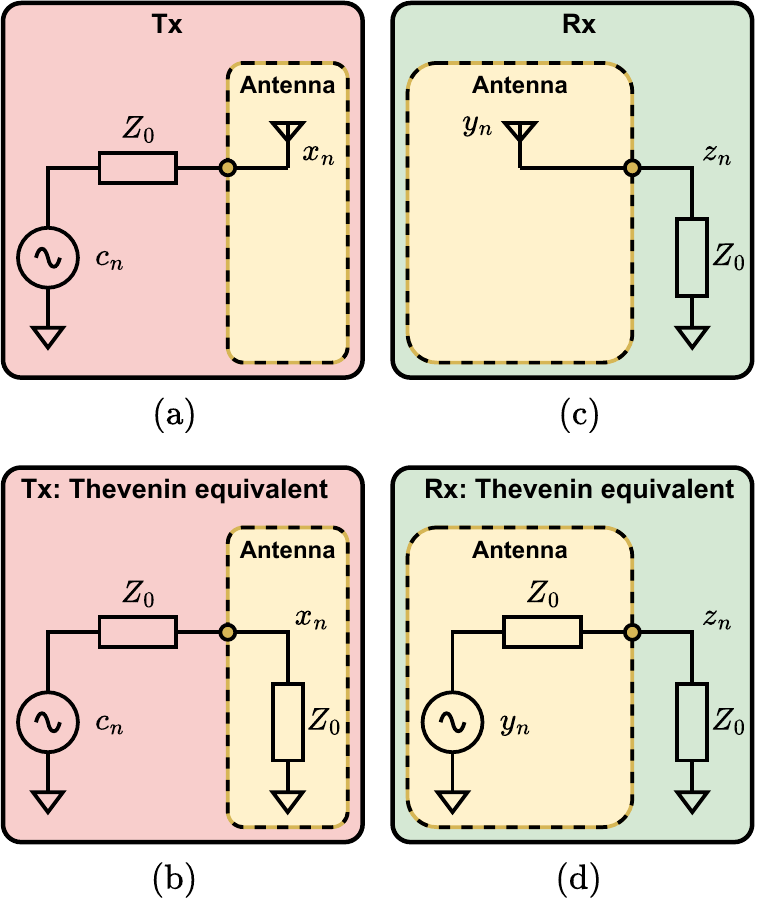}
\caption{Single-antenna transmitter and receiver in (a) and (c), respectively, and their representation using the antenna Thevenin equivalent circuit in (b) and (d), respectively.}
\label{fig:Thevenin}
\end{figure}

In this appendix, we clarify the relationship between the signal at the transmitting/receiving \gls{rf} chains and the signal at the transmitting/receiving antennas in a system operating digital beamforming.
We first consider the transmitter side and then show that a similar discussion also holds at the receiver.

In a transmitter operating digital beamforming, an \gls{rf} chain is connected to each transmitting antenna.
The circuit at the $n$th transmitting antenna is reported in Fig.~\ref{fig:Thevenin}(a), where the transmitting \gls{rf} chain is represented through a voltage generator with its series impedance $Z_0$, typically $Z_0=50~\Omega$.
In Fig.~\ref{fig:Thevenin}(a), the signal given in the output of the $n$th \gls{rf} chain is denoted as $c_n\in\mathbb{C}$ and the signal at the $n$th transmitting antenna is $x_n\in\mathbb{C}$.
To understand the relationship between $c_n$ and $x_n$, we represent the $n$th transmitting antenna through its Thevenin equivalent circuit in Fig.~\ref{fig:Thevenin}(b) according to antenna theory \cite[Chapter~2.13]{bal15}.
Assuming the $n$th transmitting antenna to be perfectly matched to $Z_0$, its input impedance is $Z_0$ and it can be equivalently represented as an impedance $Z_0$ connected to ground \cite[Chapter~2.13]{bal15}.
Thus, following the voltage divider in Fig.~\ref{fig:Thevenin}(b), we readily obtain that $x_n=c_n/2$ at the $n$th antenna of the transmitter, giving $\mathbf{x}=\mathbf{c}/2$, where $\mathbf{x}=[x_1,\ldots,x_{N_T}]$ and $\mathbf{c}=[c_1,\ldots,c_{N_T}]$.

Also at the receiver side, an \gls{rf} chain is connected to each receiving antenna in the case of digital beamforming.
The circuit at the $n$th receiving antenna is reported in Fig.~\ref{fig:Thevenin}(c), where the receiving \gls{rf} chain is assumed to be perfectly matched to $Z_0$ and thus is represented through a load $Z_0$ connected to ground.
In Fig.~\ref{fig:Thevenin}(c), the signal read in the input of the $n$th \gls{rf} chain is $z_n\in\mathbb{C}$ and the signal at the $n$th receiving antenna is $y_n\in\mathbb{C}$.
To grasp the relationship between $z_n$ and $y_n$, the $n$th receiving antenna can be represented through its Thevenin equivalent circuit, as shown in Fig.~\ref{fig:Thevenin}(d) following antenna theory \cite[Chapter~2.13]{bal15}.
In detail, assuming the $n$th receiving antenna to be perfectly matched to $Z_0$, it can be equivalently represented as a source generator imposing the voltage $y_n$ with its series impedance $Z_0$ \cite[Chapter~2.13]{bal15}.
Thus, from the voltage divider in Fig.~\ref{fig:Thevenin}(d), we derive that $z_n=y_n/2$ at the $n$th antenna of the receiver, giving $\mathbf{z}=\mathbf{y}/2$, where $\mathbf{z}=[z_1,\ldots,z_{N_R}]$ and $\mathbf{y}=[y_1,\ldots,y_{N_R}]$.

\bibliographystyle{IEEEtran}
\bibliography{IEEEabrv,main}

\end{document}